\documentclass[11pt,a4paper]{article}
\pdfoutput=1
\usepackage{jcappub,graphicx,subfigure,enumerate,latexsym}
\usepackage[english]{babel}

\providecommand{\dif}{\mathrm{d}} 
\def\beq{\begin{equation}}
\def\eeq{\end{equation}}
\def\bea{\begin{eqnarray}}
\def\eea{\end{eqnarray}}
\newcommand{\RN}{Reissner\discretionary{--}{--}{--}Nordstr\"om}
\newcommand{\KN}{Kerr\discretionary{--}{--}{--}Newman}

\def\d{\dif}
\def\lightarrow{\rightarrow}

\def\tid{b} 

\def\xx{x} \def\yy{y} \def\JJ{J} \def\EE{E} \def\rr{r} \def\tt{\theta} \def\bb{\tid}

\def\p{P}
\def\x{X}
\def\dr{\delta r}
\def\dt{\delta \theta}
\def\Veff{V_{\rm eff}}
\def\af{\zeta}
\def\hp{\hat{P}}
\def\hx{\hat{X}}

\begin{document}

\title{String loops in the field of braneworld spherically symmetric black holes and naked sigularities}
\author{Z. Stuchl\'{\i}k}
\author{M. Kolo\v{s}}
\affiliation{Institute of Physics, Faculty of Philosophy \& Science, Silesian University in Opava, Bezru\v{c}ovo n\'{a}m.13, CZ-74601 Opava, Czech Republic}
\emailAdd{zdenek.stuchlik@fpf.slu.cz}
\emailAdd{martin.kolos@fpf.slu.cz}

\abstract{
We study motion of current-carrying string loops in the field of braneworld spherically symmetric black holes and naked singularities. The spacetime is described by the \RN{} geometry with tidal charge $b$ reflecting the non-local tidal effects coming from the external dimension; both positive and negative values of the spacetime parameter $b$ are considered. We restrict attention to the axisymmetric motion of string loops when the motion can be fully governed by an appropriately defined effective potential related to the energy and angular momentum of the string loops. In dependence on these two constants of the motion, the string loops can be captured, trapped, or can escape to infinity. In close vicinity of stable equilibrium points at the centre of trapped states the motion is regular. We describe how it is transformed to chaotic motion with growing energy of the string loop. In the field of naked singularities the trapped states located off the equatorial plane of the system exist and trajectories unable to cross the equatorial plane occur, contrary to the trajectories in the field of black holes where crossing the equatorial plane is always admitted. We concentrate our attention to the so called transmutation effect when the string loops are accelerated in the deep gravitational field near the black hole or naked singularity by transforming the oscillatory energy to the energy of the transitional motion. We demonstrate that the influence of the tidal charge can be substantial especially in the naked singularity spacetimes with $b>1$ where the acceleration to ultrarelativistic velocities with Lorentz factor $\gamma \sim 100$ can be reached, being more than one order higher in comparison with those obtained in the black hole spacetimes. 
}


  
\maketitle

\section{Introduction}\label{intro}

The braneworld models of the Universe \citep{Ran-Sun:1999:PHYSR4:,Ran-Sun:1999:PHYSR4:2:} represent a five-dimensional spacetime (bulk), where the 3D brane - our Universe - is placed. The so called Randall and Sundrum (RS II) model \cite{Ran-Sun:1999:PHYSR4:} allows gravity localized near the brane with an infinite size extra dimension while the warped spacetime satisfies the 5D Einstein equations with negative cosmological constant. An arbitrary energy-momentum tensor is allowed on the brane and the effective 4D Einstein equations have to be fulfilled on the brane \cite{Maar:2004:Liv.Rev.:}. The RS II model implies the standard 4D Einstein equations in the low energy limit, but significant deviations occur at high energies, near black holes, naked singularities, or compact stars. The combined effect of the high-energy (local) and bulk (non-local) stresses alters the matching problem on the brane in comparison with the standard 4D gravity \cite{Ger-Maa:2001:PRD:}. The bulk gravity stresses imply that the matching conditions do not have unique solution on the brane. 

No exact solution of 5D braneworld Einstein equations is known at present, but numerical solutions have been found recently \cite{Fig-Wis:2011:ArXiv:}. 4D stationary and axisymmetric vacuum solution describing a braneworld rotating black hole has been found by solving the braneworld constrained 4D effective Einstein equations \cite{Ali-Gum:2005:PHYSR4:}. Of course, it is not an exact solution satisfying the full system of 5D equations, but in the framework of the constrained equations it represents a consistent rotating black hole solution reflecting the influence of the extra dimension through a single braneworld parameter. The braneworld rotating black holes are described by the metric tensor of the Kerr-Newman form with the braneworld tidal charge $b$ describing the 5D non-local gravitational coupling between the brane and the bulk \cite{Ali-Gum:2005:PHYSR4:}. For non-rotating braneworld black holes, the metric is reduced to the 
Reissner-Nordstr$\mathrm{\ddot o}$m type form containing the tidal charge \cite{Dad-Maa-Pap-Rez:2000:PHLET2:}; note that this spacetime can also represent the external field of braneworld neutron stars approximated by the uniform density internal spacetime \cite{Ger-Maa:2001:PRD:}. The braneworld tidal charge can be, in principle, both positive and negative, but the negative values are probably more relevant \cite{Dad-Maa-Pap-Rez:2000:PHLET2:}. Notice that for $b>0$ the braneworld spacetime can be identified with the \KN{} (\RN{}) spacetime by $b \rightarrow Q^2\,$, where $Q^2$ is the squared electric charge; however it is not the \KN{} (\RN{}) background since the electromagnetic part of this background is missing. Therefore, the analysis of uncharged particle motion in the standard Kerr-Newman (\RN{}) spacetimes is relevant for the properties of the charged particle motion in the braneworld rotating black holes with $b>0$. On the other hand, the results related to $b>0$ are relevant also in the case of the standard \KN{} (\RN{}) spacetimes (with $b \rightarrow Q^2$), if related to the motion of uncharged particles \cite{Stu-Hle:2002:ActaPhysSlov:,Pug-etal:2011:PHYSR4:,Oli-etal:2011:MPLA:} and uncharged strings. 

Influence of the braneworld tidal charge $b$ on the motion of test particles and related physical phenomena in vicinity of compact objects has been extensively investigated for both the black holes \cite{Stu-Kot:2009:GRG:,Ali-Tal:2009:PRD:,Sche-Stu:2009:GRG:,Sche-Stu:2009:IJMPD:,Abd-Ahm:2010:PRD:,Bin-Nun:2010a:PHYSR4:,Bin-Nun:2010b:PHYSR4:,Stu-Bla-Sla:2011:CLAQG:,Ali-etal:2012:ArXiv:} and neutron stars \cite{Kot-Stu-Tor:2008:CLAQG:,Mam-Hak-Toj:2010:MPLA:,Mor-Ahm-Abd-Mam:2010:ASS:,Mor-Ahm:2010:ArXiv:,Stu-Hla-Urb:2011:GRG:,Hla-Stu:2011:JCAP:}, or in the weak field limit \cite{Boh-etal:2008:CLAQG:,Boh-etal:2010:CLAQG:,Hor-Ger:2012:ArXiv:,Vir-Ell:2002:PhRvD:}. Here we extend mapping of the properties of the braneworld black hole or naked singularity spacetimes to the case of the axisymmetric motion of current-carrying string loops that could in a simplified way represent plasma exhibiting string-like behavior via dynamics of the magnetic force lines in the plasma \cite{Sem-Ber:1990:ASS:,Chri-Hin:1999:PhRvD:,Sem-Dya-Pun:2004:Sci:} or due to thin isolated flux tubes of plasma that could be described by an one-dimensional string \cite{Spr:1981:AA:}; the string loops were also studied in \citep{Lar:1994:CLAQG:,Fro-Lar:1999:CLAQG:,Gu-Cheng:2007:GRG:,Wang-Cheng:2012:ArXiv:,Iga-Ish:2010:PHYSR4:}. Tension of the string loops prevents their expansion beyond some radius, while their worldsheet current introduces an angular momentum barrier preventing the loop from collapsing into the black hole or naked singularity. Naturally, the mapping is relevant also for oscillatory motion of cosmic string loops. The string-like phenomena could be relevant in oscillatory effects related to the high frequency quasiperiodic oscillation in accretion discs orbiting black holes - see, e.g., \cite{Tor-etal:2005:ASTRA:}.

Relativistic current-carrying strings moving axisymmetrically along the axis of a Kerr black hole has been studied in \cite{Jac-Sot:2009:PHYSR4:} where it has been proposed that such a string loop configuration can be used as a model of jet formation and acceleration in the field of black holes in microquasars or active galactic nuclei. The role of the cosmic repulsion in the string loop motion has been investigated for the Schwarzschild---de Sitter (SdS) black hole spacetime in \citep{Kol-Stu:2010:PHYSR4:}, and the acceleration of the strings in the SdS spacetime has been discussed in \cite{Stu-Kol:2012:PHYSR4:}. It has been demonstrated that a crucial role plays the so called static radius of the SdS spacetimes \cite{Stu-Hle:1999:PHYSR4:} where the gravitational attraction of the black hole is just balanced by the cosmic repulsion; note that the static radius is important also for the test particle motion \cite{Stu-Kov:2008:IJMPD:,Stu-Sche:2011:JCAP:} and the perfect fluid toroidal configurations \cite{Stu-Sla-Hle:2000:ASTRA:}.

Here we shall test the role of the tidal charge $b$ of the braneworld black hole and naked singularity spacetimes in the motion of string loops. The string loop is assumed to be threaded onto an axis of the \RN{} spacetime that is chosen to be the y-axis as demonstrated by Fig. 1 in \cite{Kol-Stu:2010:PHYSR4:}. The string loop oscillates in the $x$-$z$ plane propagating simultaneously in the $y$-direction. Due to the assumed axisymmetry of the string motion one point path can represent whole movement of the string. Trajectory of the string can be represented by a curve in the 2D $x$-$y$ plane. 

In dependence on the tidal parameter $b$ we determine the regions of the parameter space allowing for the capturing, trapping and escaping of the string loops. We analyze transition between the regular and chaotic character of the string loop motion. Finally, we focus our attention to the possibility to transmute the motion of a string loop originally oscillating around a black hole (naked singularity) in one direction to the perpendicular direction, modeling thus an accelerating jet. Due to the chaotic character of the string loop motion such a transformation of the energy from the oscillatory to the transitional mode is possible \cite{Lar:1994:CLAQG:,Jac-Sot:2009:PHYSR4:,Kol-Stu:2010:PHYSR4:}. We demonstrate that the role of the braneworld tidal parameter is significant, especially in the case of the RN naked singularity spacetimes when the string can be accelerated up to ultrarelativistic velocities. 
\section{Current-carrying string loop motion in spherically symmetric spacetimes}

The string loop motion is governed by barriers given by the string tension and the worldsheet current determining the angular momentum. These barriers are modified by the gravitational field of a braneworld RN black hole or naked singularity characterized by the mass $M$ and tidal charge $b$ \cite{Stu-Kot:2009:GRG:}. For the spherically symmetric spacetimes the line element expressed in the standard Schwarzschild coordinates and the geometric units ($c=G=1$) reads
\beq
 \d s^2 = A(r) \d t^2 + A^{-1}(r) \d r^2 + r^2 (\d \theta^2 + \sin^2 \d \phi^2), \label{SfSymMetrika}
\eeq
where the characteristic function takes the form 
\beq
A(r) = \left(1-\frac{2M}{r} + \frac{\tid}{r^2} \right). \label{TCHmetrika}
\eeq
In our study we use also the Cartesian coordinates
\beq
 x = r \sin(\theta),\quad y = r \cos(\theta). \label{ccord}
\eeq

The string worldsheet is described by the spacetime coordinates $X^{\alpha}(\sigma^{a})$ (with $\alpha = 0,1,2,3$) considered as functions of two worldsheet coordinates $\sigma^{a}$ (with $a = 0,1$). We use worldsheet coordinates $(\tau, \sigma)$; $\tau$ denotes some affine parameter related to the proper time measured along the moving string, $\sigma$ reflects the axial symmetry of the oscillating string. Dynamics of the string loops is described by the action related to the string tension $\mu > 0$ and a scalar field $\varphi$; $ \varphi_{,a} = j_a $ determines current (angular momentum) of the string \cite{Jac-Sot:2009:PHYSR4:}. The assumption of axisymmetry of the string loops implies the scalar field in linear form with constants $j_{\sigma}$ and 
$j_{\tau}$  
\beq
   \varphi = j_{\sigma}\sigma + j_{\tau}\tau.
\eeq

Integration of the equations of motion of asymmetric string loops or open strings is a very complex task that has to be treated by numerical methods only \cite{Gar-Will:1987:PHYSR4:,Vil-She:1994:CSTD:,Lar:1994:CLAQG:,Fro-Lar:1999:CLAQG:,Pog-Vach:1999:PHYSR4:,Pag:1999:PHYSR4:,DeV-Fro:1999:CLAQG:,Sna-Fro-DeV:2002:CLAQG:,Fro-Fur:2001:PHYSR4:,Sna-Fro:2003:CLAQG:,Fro-Ste:2004:PHYSR4:}. The symmetry of the braneworld \RN{} spacetime and the assumption of the axisymmetric string oscillations enables a substantial simplification - the string motion can be treated using the Hamiltonian formalism related to simple, one-particle motion. As demonstrated in \citep{Lar:1993:CLAQG:,Stu-Kol:2012:PHYSR4:}, the string loop motion in spherically symmetric spacetimes can be described by the Hamiltonian 
\beq
 H = \frac{1}{2} g^{\alpha\beta} \p_\alpha \p_\beta + \frac{1}{2} \mu^2 r^2 \sin^2\,\theta + \mu J^2 + \frac{1}{2} \frac{(j_\tau^2 - j_\sigma^2)^2}{r^2 \sin^2\,\theta}, 
\eeq
where $\p_{\alpha}$ is 4-momentum and $\alpha, \beta$ correspond to coordinates $t,r,\theta,\phi$. For axisymmetric oscillations of string loops in the spherically symmetric spacetimes we can use two constants of motion -  string energy $E$ and total angular momentum squared $J^2$ given by the relations (for details see \cite{Stu-Kol:2012:PHYSR4:})  
\beq
            E = - \p_t ,   \quad  J^2 = j_{\sigma}^2 + j_{\tau}^2. 
\eeq
The spacetimes symmetries imply existence of the other constant of motion
\beq
                \p_\phi = L = -2 j_\tau j_\sigma  \label{SAngMomentum}
\eeq
that do not enter the equations of motion due to the spherical symmetry of the spacetime. Then the Hamiltonian takes the form 
\beq
 H = \frac{1}{2} A(r) \p_r^2 + \frac{1}{2r^2} \p_\theta^2 - \frac{E^2}{2A(r)}  + \frac{\Veff(r,\theta)}{A(r)}. \label{HamHam}
\eeq
The effective potential $\Veff(r,\theta)$ for the string loop motion is given by the relation
\beq
\Veff(r,\theta) = \frac{1}{2} A(r) \left( \mu \, r\sin\theta  + \frac{J^2}{r\sin\theta} \right)^2.
\eeq

Using an affine parameter of the string motion $\af$, the Hamilton–-Jacobi equations
\beq
 \frac{\d \x^\mu}{\d \af} = \frac{\partial H}{\partial \p_\mu}, \quad
 \frac{\d \p_\mu}{\d \af} = - \frac{\partial H}{\partial \x^\mu} \label{Ham_eq}
\eeq
imply equations of the string loop motion in the form
\bea
 \dot{r} = A \p_r, \quad \dot{\p_r} &=& \frac{1}{A} \frac{\p_\theta^2}{r^4} 
            \left( A r - \frac{1}{2} \frac{\d A}{\d r}   r^2  \right)
            - \frac{\d A}{\d r} \p_r^2  - \frac{1}{A}\frac{\d \Veff}{\d r}, \label{EqOfMotionB1} \\
 \dot{\theta} = \frac{\p_\theta}{r^2},  \,\,\,\quad  \dot{\p_\theta} &=&  - \frac{1}{A}\frac{\d \Veff}{\d \theta}. \label{EqOfMotionB2}
\eea
Dot means derivative with respect to the affine parameter: $\dot{f} = \d f/\d \af$. 

The energy of the string loop can be expressed in the form \citep{Lar:1993:CLAQG:,Stu-Kol:2012:PHYSR4:}
\beq
 E^2 = \dot{r}^{2} + A(r) r^2 \dot{\theta}^{2} + 2 \Veff. \label{StringEnergy}
\eeq
The loci where the string loop has zero velocity ($\dot{r}=0, \dot{\theta}=0$) forms boundary of the string motion. The boundary energy function can be defined by the relation 
\beq
                 E^2_{\rm b}(x,y) = 2 \Veff
\eeq
that is the rescaled effective potential.
The string loop motion is confined to the region where $\Veff(r,\theta) \leq 0$.

The boundary energy function of the string loop motion expressed in the Cartesian coordinates reads
\beq
E_{\rm b}(x,y) = \sqrt{A(r)}\left( \frac{J^2}{x} + x \mu \right) = \sqrt{A(r)} f(x), \label{EqEbXY}
\eeq
where $r = r(x,y) = \sqrt{x^2+y^2}$. The function $A(r)$ reflects the spacetime properties, $f(x)$ those of the string loop. The behavior of the boundary energy function is given by the interplay of the functions $A(r)$ and $f(x)$. Assuming that the string loop will start its motion from rest, i.e., assuming $\dot{r}(0) = 0$ and $\dot{\theta}(0)=0$, the initial position of the string loop will be located at some point of the energy boundary function $E_{\rm b}(x,y)$ of its motion.

Stationary points of $E_{\rm b}(x,y)$ are determined by the conditions
\bea
  (E_{\rm b})_{x}' &=& 0  \,\, \ \Leftrightarrow \,\,  x A_{r}' f = - 2 r A f_{x}' \label{extr_a1} \\
  (E_{\rm b})_{y}' &=& 0  \,\, \ \Leftrightarrow \,\,   A_{r}' y  = 0, \label{extr_a2}
\eea
where we assume $f(x) > 0$ for $x>0$. The prime $()_{m}'$ denotes derivation with respect to the coordinate $m$.
In order to determine character of the stationary points at $(x_{\rm e},y_{\rm e})$ given by the stationarity conditions, i.e., whether we have a maximum ("hill") or minimum ("valley") of the energy boundary function $E_{\rm b}(x,y)$, we have to study the conditions
\bea
 && [(E_{\rm b})_{yy}''] (x_{\rm e},y_{\rm e}) \quad < 0 \,\, \mathrm{(max)} \quad > 0 \,\, \mathrm{(min)}  \label{extr_b1}\\
 && [(E_{\rm b})_{yy}'' (E_{\rm b})_{xx}'' - (E_{\rm b})_{yx}'' (E_{\rm b})_{xy}''](x_{\rm e},y_{\rm e}) > 0 . \label{extr_b2}
\eea
The behavior of the energy boundary function $E_{\rm b}(x,y)$ in the $x$-direction in the equatorial plane ($y=0$) determines opening of the string loop motion to the origin of the coordinate system (black hole horizon), while its behavior in the $y$-direction determines opening of the motion to infinity. 

It is obvious from the equation (\ref{EqEbXY}) that we can make the rescaling $E_{\rm b} \rightarrow E_{\rm b} / \mu $ and $J \rightarrow J / \sqrt{\mu} $, assuming $\mu > 0$. This choice of ``units'' will not affect the character of the string boundary energy function, and it is equivalent to the assumption of the string tension $\mu=1$ in Eq (\ref{EqEbXY}) - see \cite{Kol-Stu:2010:PHYSR4:,Stu-Kol:2012:PHYSR4:}. In the following, we use this simplification.

\section{Analysis of the string loop motion in braneworld RN metric}
\subsection{The braneworld spacetimes}
The spherically symmetric braneworld \RN{} metric with the line element (\ref{SfSymMetrika}) and the characteristic function $A(r)$ describes a black hole when the spacetime parameters fulfill the inequality 
\beq
              M^2 \geq \tid ;
\eeq
otherwise the spacetime describes a naked singularity. Clearly, the naked singularity spacetimes can occur only for positive $\tid$. In this case the metric (\ref{TCHmetrika}) describes also the standard \RN{} background where 
$\tid \lightarrow Q^2 > 0$; however, in such a case there is also the electromagnetic field of Coulombic character present in the RN spacetime \cite{Mis-Tho-Whe:1973:Gra:}. For negative values of the tidal charge ($\tid  < 0$) we can have only braneworld black holes. Detailed discussion of the properties of the braneworld \RN{} spacetimes and the test particle motion in this spacetime can be found in \cite{Kot-Stu-Tor:2008:CLAQG:}. 

In order to discuss properties of the string loop motion, we introduce dimensionless coordinates and motion constants $\xx \rightarrow x/M, \yy \rightarrow y/M, \rr \rightarrow r/M, \bb \rightarrow \tid /M^2, \JJ \rightarrow J/M, \EE \rightarrow E/M$. We can assume $\JJ>0$ due to the spherical symmetry of the spacetime, similarly to the case of the motion of test particles. The detailed discussion of the properties of the effective potential and the string loop motion in the spherically symmetric Schwarzschild spacetime can be found in \cite{Kol-Stu:2010:PHYSR4:}. Here we extend the discussion to  the case of the behavor of the boundary energy function in the braneworld \RN{} spacetimes. 

The geometry (\ref{TCHmetrika}) introduces a characteristic length scale corresponding to the radius of the black hole horizon that is given by the condition $A(r)=0$. The boundary energy function $\EE_{\rm b}$ (\ref{EqEbXY}) is defined in the so called static region where $A(r)>0$. Its boundary ($A(r)=0$) gives the loci of the event horizons that are determined by the condition 
\beq
 \bb = \bb_{\rm hor}(r) \equiv 2r - r^2.
\eeq
The position of the event horizons is illustrated in Fig. \ref{stringFIG_1} where the radius function is restricted to the $x$-direction without loss of generality - the so called dynamical regions where $A(r) < 0$ are shaded. We see that black holes exist for $\bb \leq 1$, while for $\bb>1$ we have naked singularity spacetimes. Two event horizons exist for $\bb>0$ while for $\bb<0$ there exist one horizon only since we consider the regions above the physical singularity at $r=0$. 

\subsection{The boundary energy function}
Now we have to discuss properties of the boundary energy function given by (\ref{EqEbXY}) in the braneworld RN spacetimes (\ref{TCHmetrika}). The energy boundary function is not defined in the dynamical region where $A(r)<0$. In the naked singularity spacetimes it is thus defined at all $r>0$. In the black hole spacetimes, we focus our attention to the region above the outer horizon ($r > r_{+}=1+(1-\tid)^{1/2}$). 
Under the simplifications introduced above, the energy boundary function takes the form 
\beq
     \EE_{\rm b}(x,y, \tid, \JJ) =  \sqrt{1-\frac{2}{r} + \frac{\tid}{r^2}}\left( x + \frac{J^2}{x} \right),
\eeq
where $r^2=x^2+y^2$. First we have to give the asymptotic behavior of (\ref{EqEbXY}). Similarly to the case of the Schwarzschild spacetime, in the $x$-direction there is 
\beq
     \EE_{\rm b}(x \lightarrow \infty, \tid, \JJ) \lightarrow +\infty , 
\eeq
while in the $y$-direction we obtain the energy boundary function related to the flat spacetime 
\beq
     \EE_{\rm b}(y \lightarrow \infty, \tid, \JJ) \lightarrow  \left( x + \frac{J^2}{x} \right) = \EE_{\rm b(flat)} ; 
\eeq
for details see \cite{Kol-Stu:2010:PHYSR4:}. Asymptotic behavior at the black hole horizon reads 
\beq 
          \EE_{\rm b}(r \lightarrow r_{+}, \tid, \JJ) \lightarrow 0 . 
\eeq
In the braneworld RN naked singularity spacetimes, allowed only for $\bb > 0$, the asymptotic behavior at the physical  singularity (at $r=0$) is important. There is
\beq 
          \EE_{\rm b}(r \lightarrow 0, \tid, \JJ) \lightarrow \infty . 
\eeq

The equations (\ref{extr_a1}-\ref{extr_a2}) determine the local extrema of the boundary energy function $\EE_{\rm b}(x,y,b,J)$:
\begin{itemize}

\item off-equatorial extrema - there are minima located at 
\beq
 \xx = \JJ, \quad \yy = \sqrt{\bb^2 - \JJ^2} ,
\eeq

\item equatorial extrema - these are determined by the relation 
\beq
               \JJ^2 = \JJ_{\rm E}^2(x;\bb) \equiv \frac{\xx^3(\xx-1)}{\xx^2 - 3\xx + 2 \bb} . \label{JexE2}
\eeq
 
\end{itemize}
Contrary to the case of the Schwarzschild spacetime \cite{Kol-Stu:2010:PHYSR4:}, the local extrema can exist off the equatorial plane. The off-equatorial extrema are only local minima in the naked singularity RN spacetimes with braneworld dimensionless tidal charge $\bb>1$ and can exist for limited angular momentum $\JJ$ restricted by the condition $\JJ < \bb$. In the equatorial plane both minima or maxima of the energy boundary function $\EE_{\rm b}(x,y;\bb,\JJ)$ can appear in both black hole and naked singularity braneworld RN spacetimes.

\begin{figure}
\includegraphics[width=0.9\hsize]{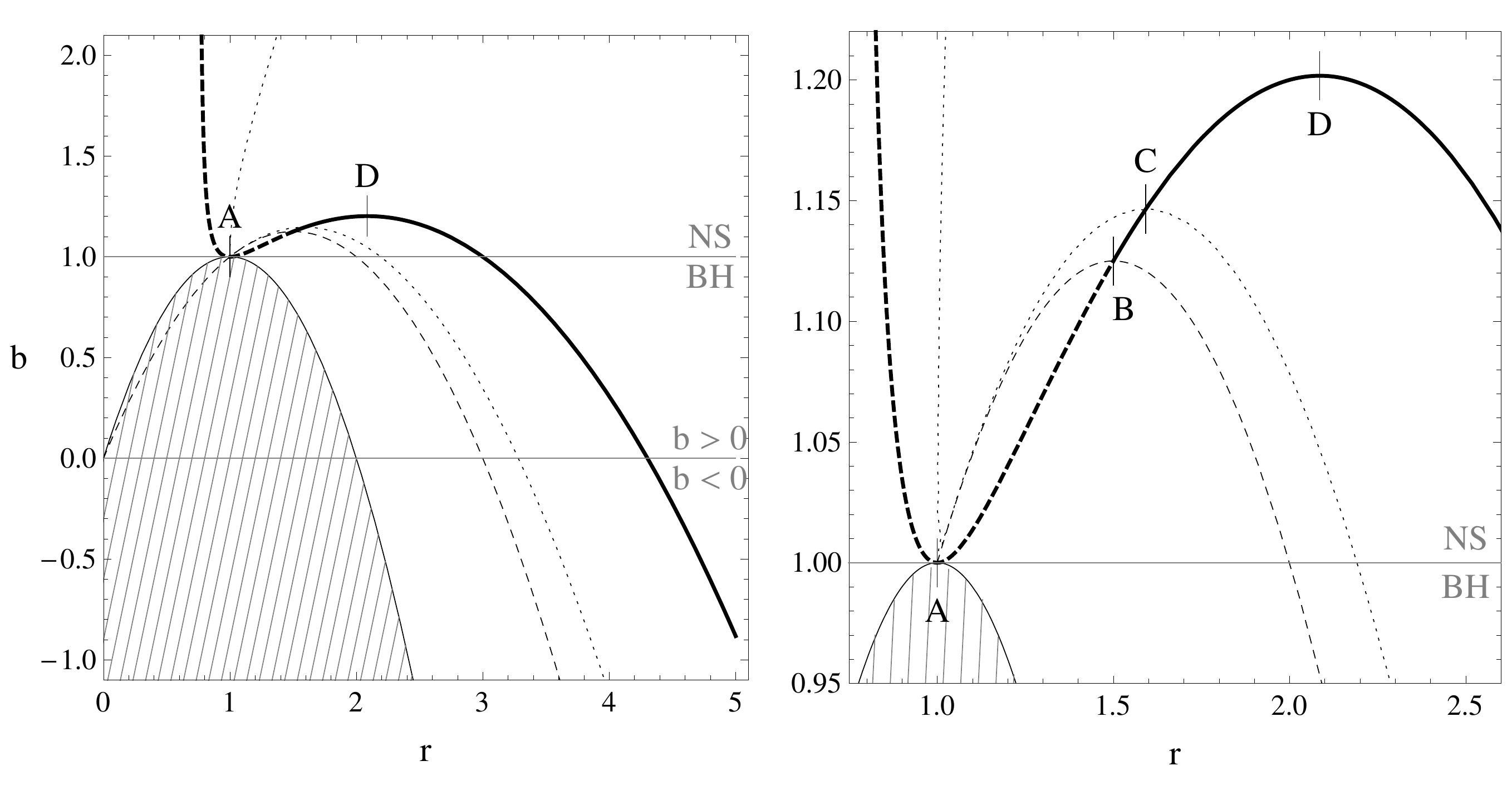}
\vspace{-0.3cm}
\caption{\label{stringFIG_1}
Characteristic tidal-charge functions determining properties of the string loop motion in the braneworld RN spacetimes in dependence on the tidal charge $\bb$. Solid curve $\bb_{\rm hor}(\xx)$ gives the position of the horizons and restricts the dynamical region (hatched). The number and position of the extrema of the function $\JJ_{\rm E}^2(\xx,\bb)$ are given by the function $\bb_{\rm ex}(\xx)$ (thick curve), the part where $\JJ_{\rm E}^2(\xx,\bb)$ is not defined is dashed. Dashed line represents the function $\bb_{\rm div}(\xx)$ - function $\JJ_{\rm E}^2(\xx,\bb)$ diverges there. Dotted curve is solution to the equation for "lake" existence $\bb_{\rm extr(L)}(x)$.
Important points have the following coordinates and significance: $A=[1,1]$, minimum of $\bb_{\rm ex}(\xx)$ , $A$ and $B=[3/2,9/8]$ are points of intersection of the curves $\bb_{\rm ex}(\xx)$ and $\bb_{\rm div}(\xx)$, $C=[1.592,1.146]$ is point of intersection of the curves $\bb_{\rm ex}(\xx)$ and $\bb_{\rm extr(L)}(x)$, $D=[2.086,1.202]$ represents the maximum of $\bb_{\rm ex}(\xx)$.}
\end{figure}

Behavior of the equatorial extrema is more complicated as compared to the off-equatorial minima and is governed by the extrema function $\JJ_{\rm E}^2(x;\bb)$. The function $\JJ_{\rm E}^{2}(x;\bb)$ diverges at radii given by the condition 
\beq
          \bb = \bb_{\rm div}(x) \equiv \frac{3\xx -\xx^2}{2} .
\eeq 
The local maximum of the function $\bb_{\rm div}(x)$ is located at 
\beq
          \xx_{\rm div(max)} = \frac{3}{2}, \quad \bb_{\rm div(max)} = \frac{9}{8} ,
\eeq
denoted as point $B = [3/2,9/8]$. The zero point of the equatorial extrema function $\JJ_{\rm E}^2(x;\bb)$ is located at 
\beq
          \xx = 1 
\eeq
independently of the braneworld tidal charge; it is relevant in all RN naked singularity spacetimes with $\bb > 1$. 
 
The local extrema of the function $\JJ_{\rm E}^2(x;\bb)$ are located at radii given by the condition
\beq
     \bb = \bb_{\rm ex}(x) \equiv \frac{\xx^3 -5\xx^2 +3\xx}{3 -4\xx}.
\eeq
This function diverges at
\beq 
          \xx_{\rm ex(div)} = \frac{3}{4} ,
\eeq
it has one physically relevant zero point at 
\beq
          \xx_{\rm ex(z)} = \frac{5 + \sqrt{13}}{2} ,
\eeq
and it has two extrema given by 
\bea
 \xx_{\rm ex(min)} &=& 1, \quad \bb_{\rm ex(min)} = 1; \\
 \xx_{\rm ex(max)} &=& \frac{3}{16} \left(7+\sqrt{17}\right) \doteq 2.086, \nonumber \\ 
 \bb_{\rm ex(max)} &=& \frac{3}{512} \left(135+17 \sqrt{17}\right) \doteq 1.202. \label{bexmax}
\eea
The local minimum (maximum) of the function $\bb_{\rm ex}(x)$ is thus at the point $A = [1,1]$ ($D = [2.086,1.202]$). 
The intersection points of the functions $\bb_{\rm ex}(x)$ and $\bb_{\rm div}(x)$ are $A = [1,1]$ and $B = [3/2,9/8]$. 

The natural limit on the extreme function, $\JJ_{\rm E}^2(x;\bb) > 0$, implies that two conditions have to be satisfied simultaneously  
\beq
           \bb \geq \bb_{\rm div}(x), \quad \xx > 1 . 
\eeq
The condition $\xx > 1$ is relevant for the naked singularity RN spacetimes only. 

\begin{figure}
\includegraphics[width=0.8\hsize]{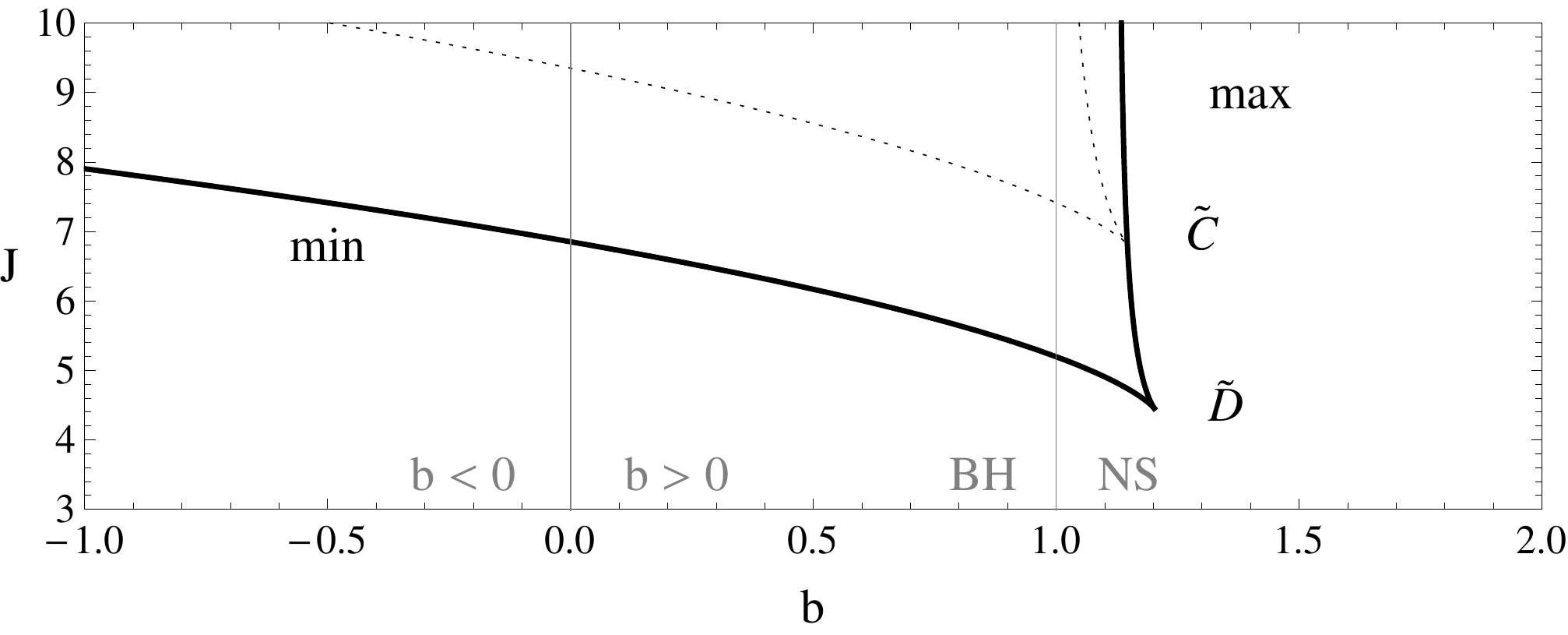}
\vspace{-0.3cm}
\caption{\label{stringFIG_2} 
Extremal values of the characteristic angular momentum functions - extremal (full lines), and "lake" (dotted lines). The points $\tilde{C}=[1.146,6.819]$ and $\tilde{D}=[1.202,4.454]$ are related to the points C and D in Fig. \ref{stringFIG_1}. }  
\end{figure}

The functions $\bb_{\rm ex}(\xx)$, $\bb_{\rm div}(\xx)$ and $\bb_{hor}(x)$ are illustrated in Fig. \ref{stringFIG_1}. The extremal values of the extremal angular momentum parameter given by the functions $\JJ_{\rm E(max)}(\bb)$ and $\JJ_{\rm E(min)}(\bb)$ are illustrated in Fig. \ref{stringFIG_2}. Recall that in the Schwarzschild spacetimes ($\bb=0$) the minimum $\JJ_{\rm E(min)}^2 = 46.936$ is located at $\xx_{\rm min} = 4.303$ \cite{Kol-Stu:2010:PHYSR4:}. 

\begin{figure}
\subfigure[~ $b=-1$ BHP]{\label{stringFIG_3a}\includegraphics[width=0.32\hsize]{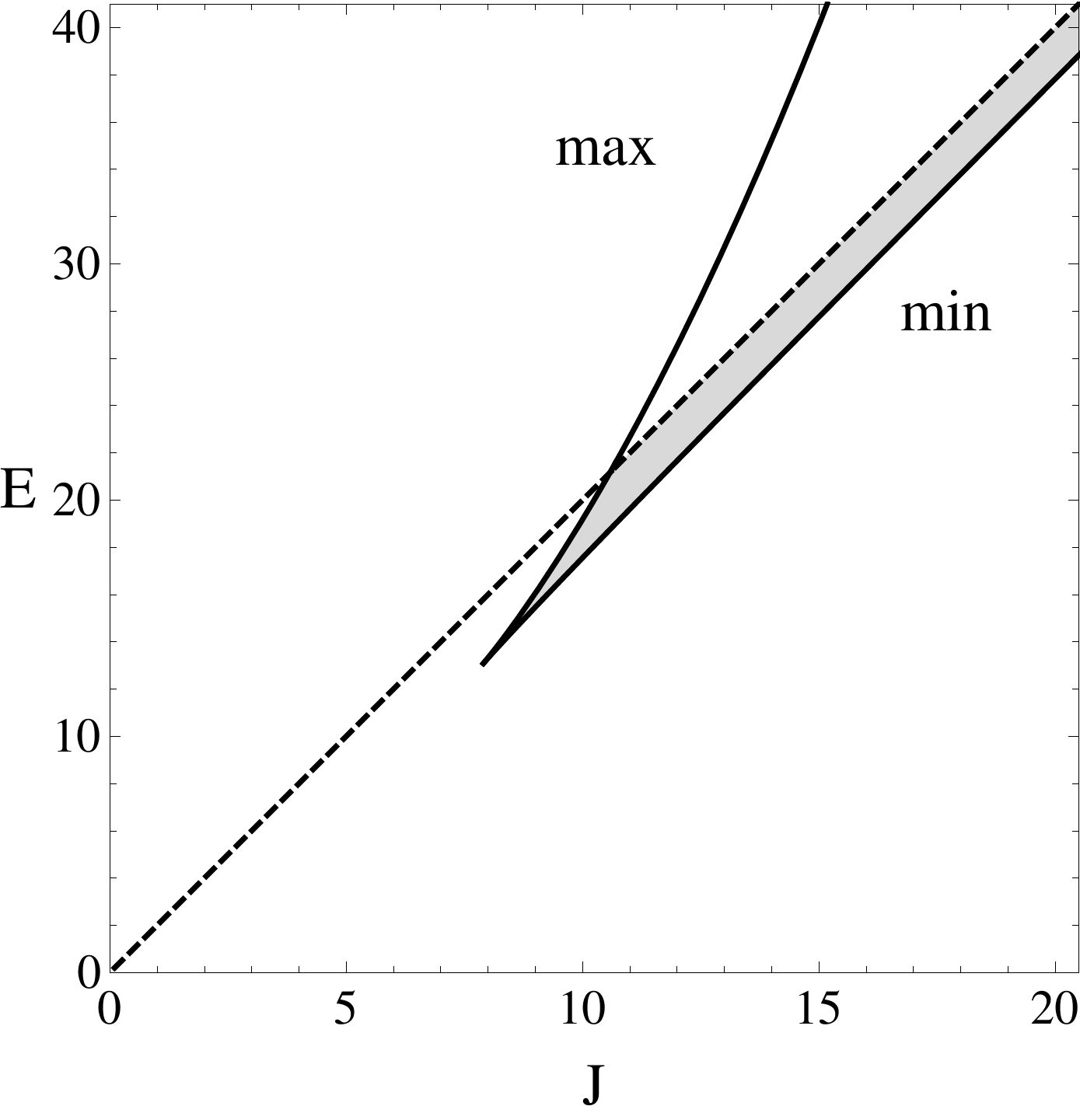}}
\subfigure[~ $b=0.99$ BHN]{\label{stringFIG_3b}\includegraphics[width=0.32\hsize]{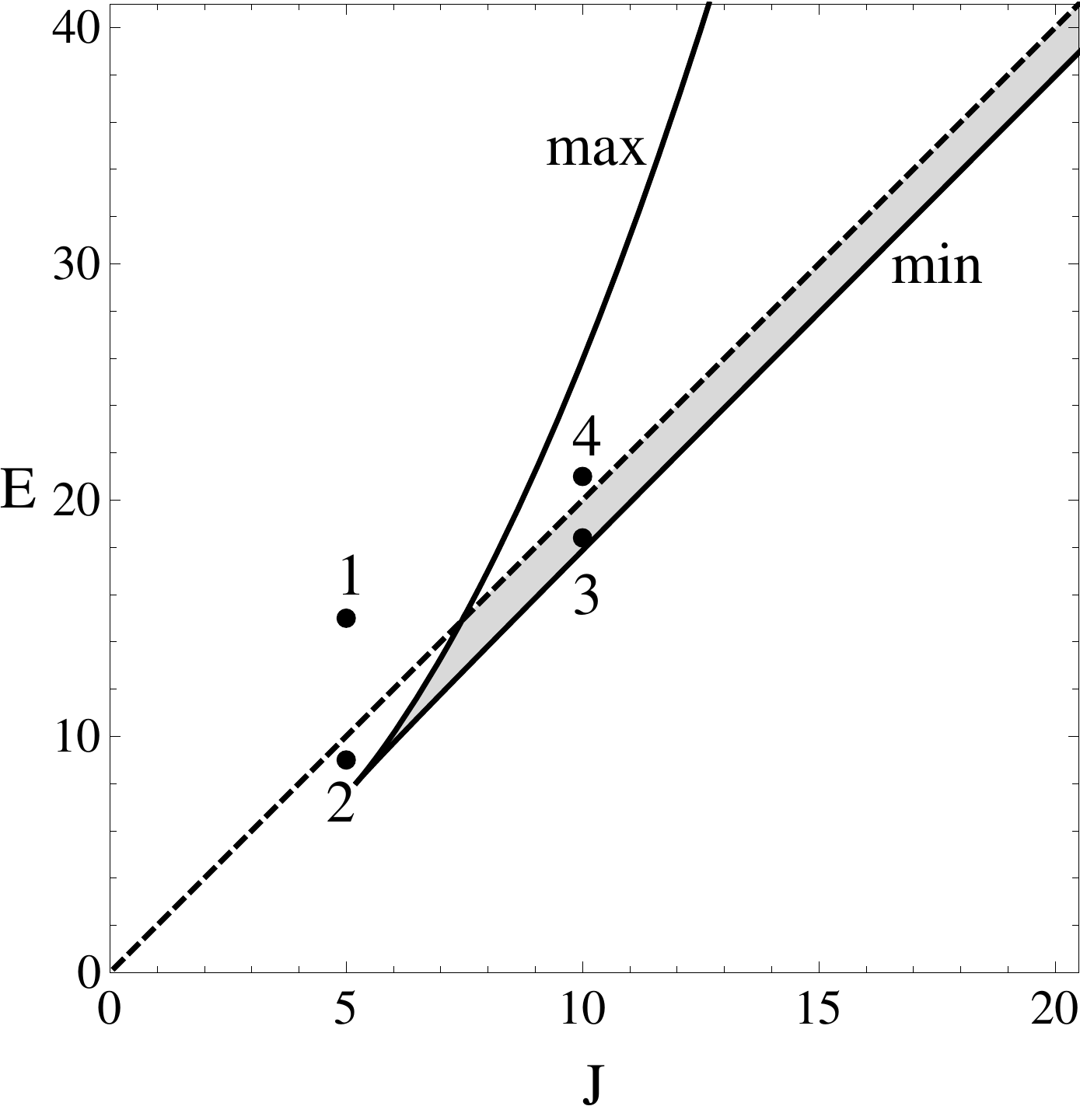}}\\
\subfigure[~ $b=1.07$ NS1]{\label{stringFIG_3c}\includegraphics[width=0.32\hsize]{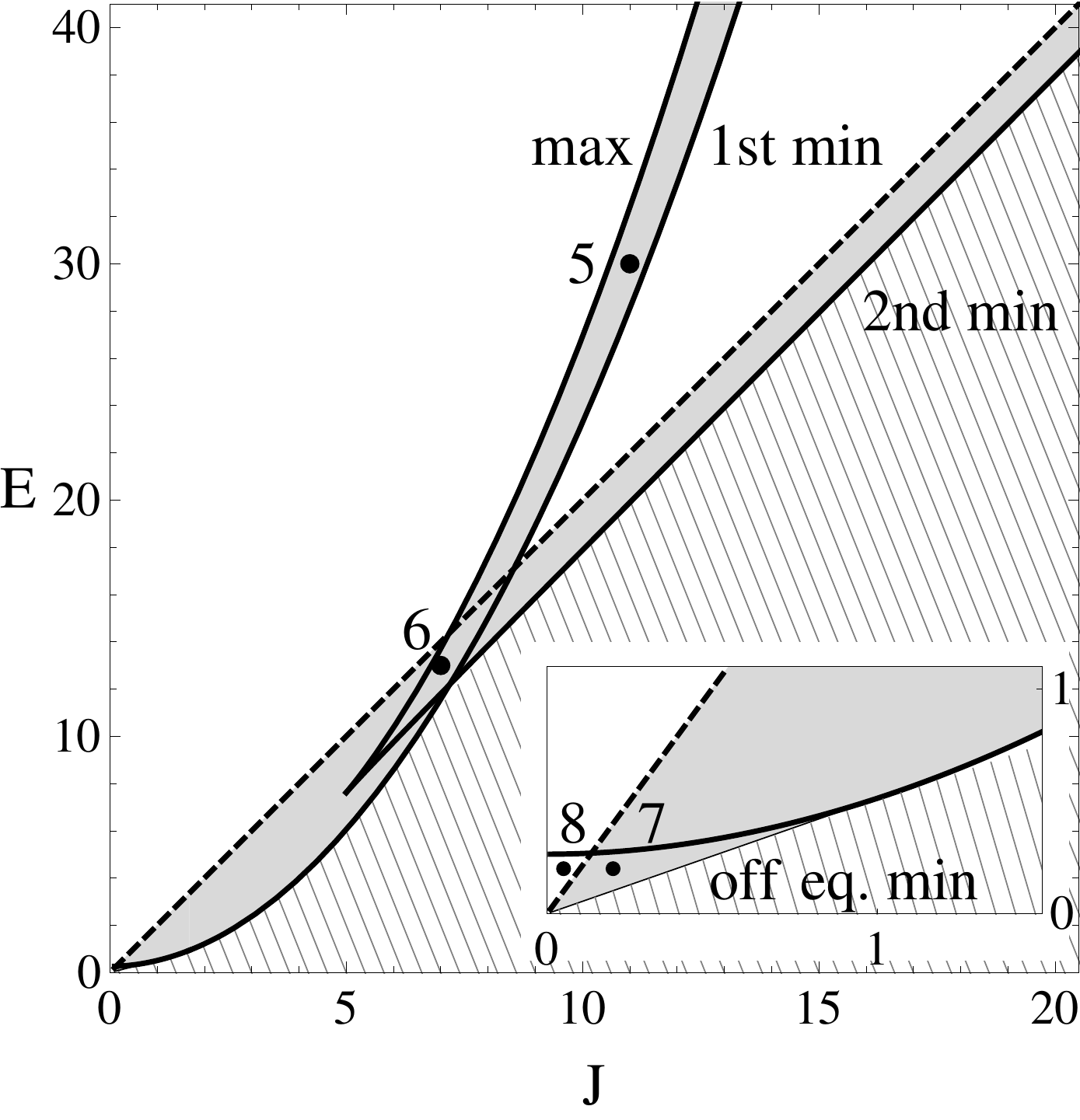}}
\subfigure[~ $b=1.14$ NS2]{\label{stringFIG_3d}\includegraphics[width=0.32\hsize]{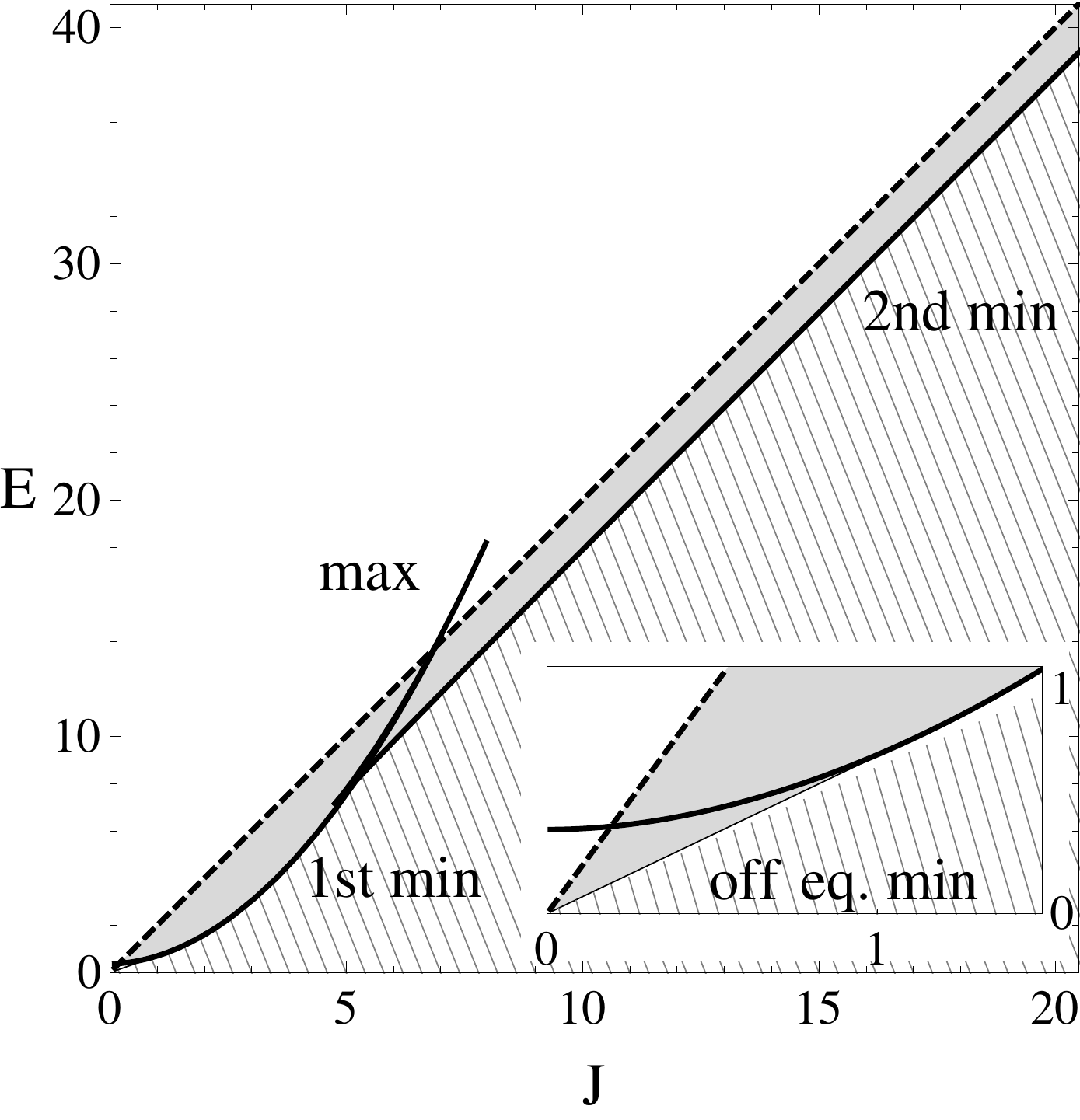}}
\subfigure[~ $b=2$ NS3]{\label{stringFIG_3e}\includegraphics[width=0.32\hsize]{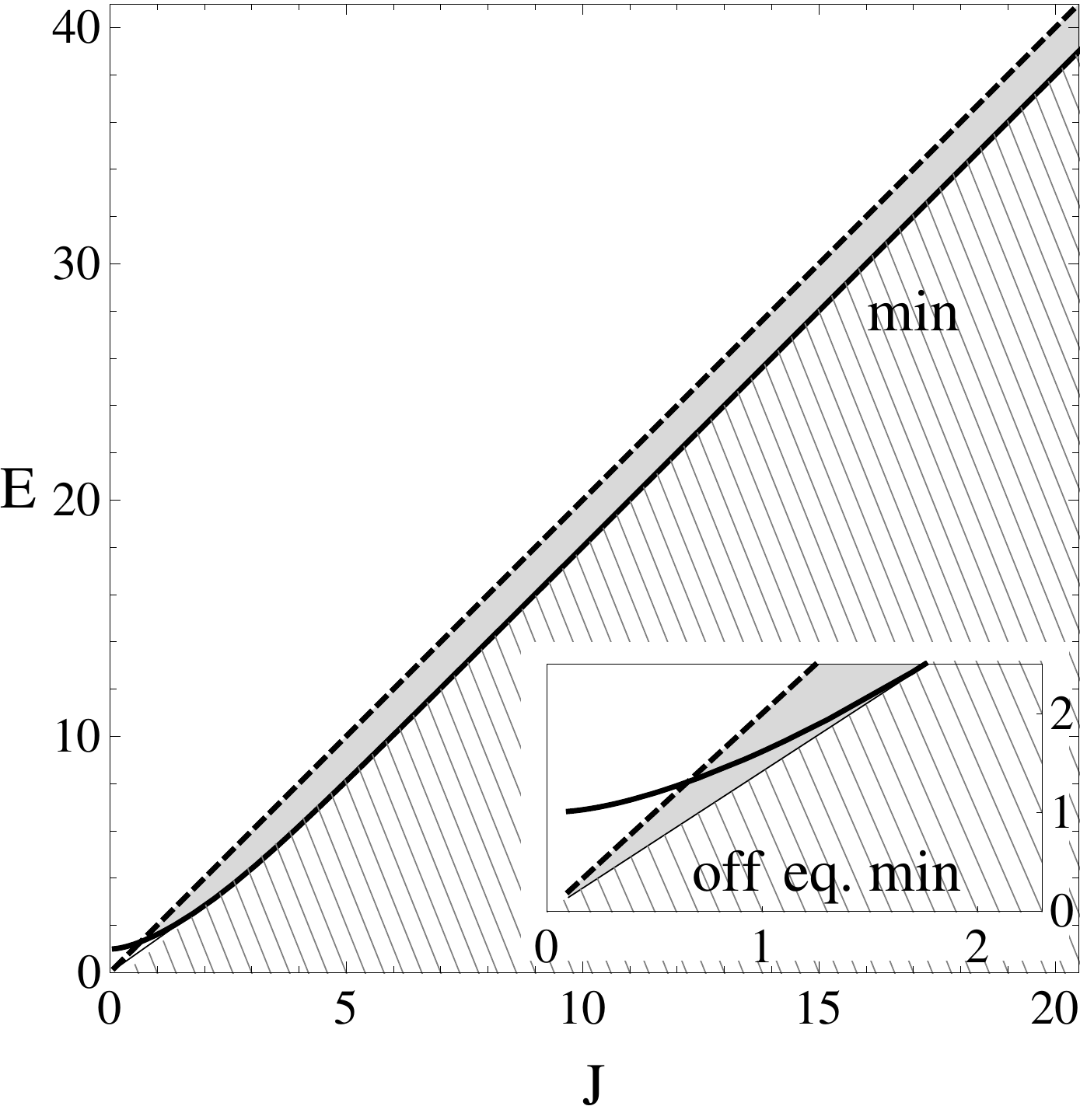}}
\vspace{-0.3cm}
\caption{\label{stringFIG_3}
Extrema of the boundary energy function $\EE_{\rm b}(\xx,\yy;\bb,\JJ)$ illustrated for the five types of the braneworld RN spacetimes; tidal charge is chosen in correspondence with Fig. \ref{stringFIG_3}. Thick solid curves correspond to the maxima and minima of the energy boundary function $\EE_{\rm b}(\xx,\yy;\bb,\JJ)$ in the equatorial plane in the $\xx$-direction. Minimum of the boundary energy function for the flat spacetime $\EE = 2 \JJ$ (escape to infinity) is represented by the dashed curve. Regions where the ``lakes'' corresponding to the trapped states can exist are shaded, regions that are not accessible for the motion are hatched. Numbers denote the examples of different types of the characteristic sections of $\EE=$~const of the boundary energy function $E_{\rm b}(\xx,\yy;\bb,\JJ)$ illustrated in Fig. \ref{stringFIG_7}.
}
\end{figure}

\begin{figure}
\subfigure[~ $b=-1$ BHN]{\label{stringFIG_4a}\includegraphics[width=0.32\hsize]{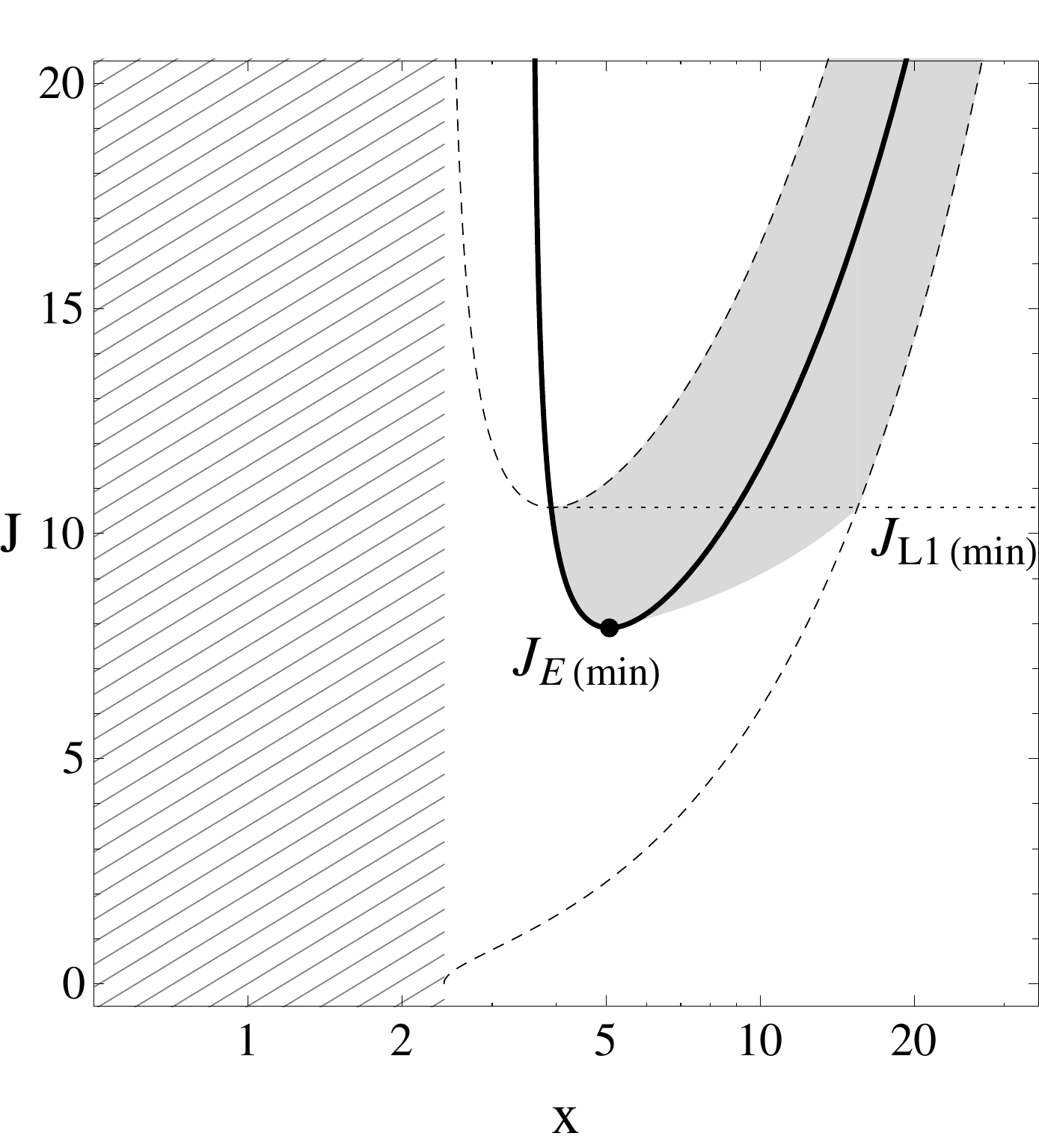}}
\subfigure[~ $b=0.99$ BHP]{\label{stringFIG_4b}\includegraphics[width=0.32\hsize]{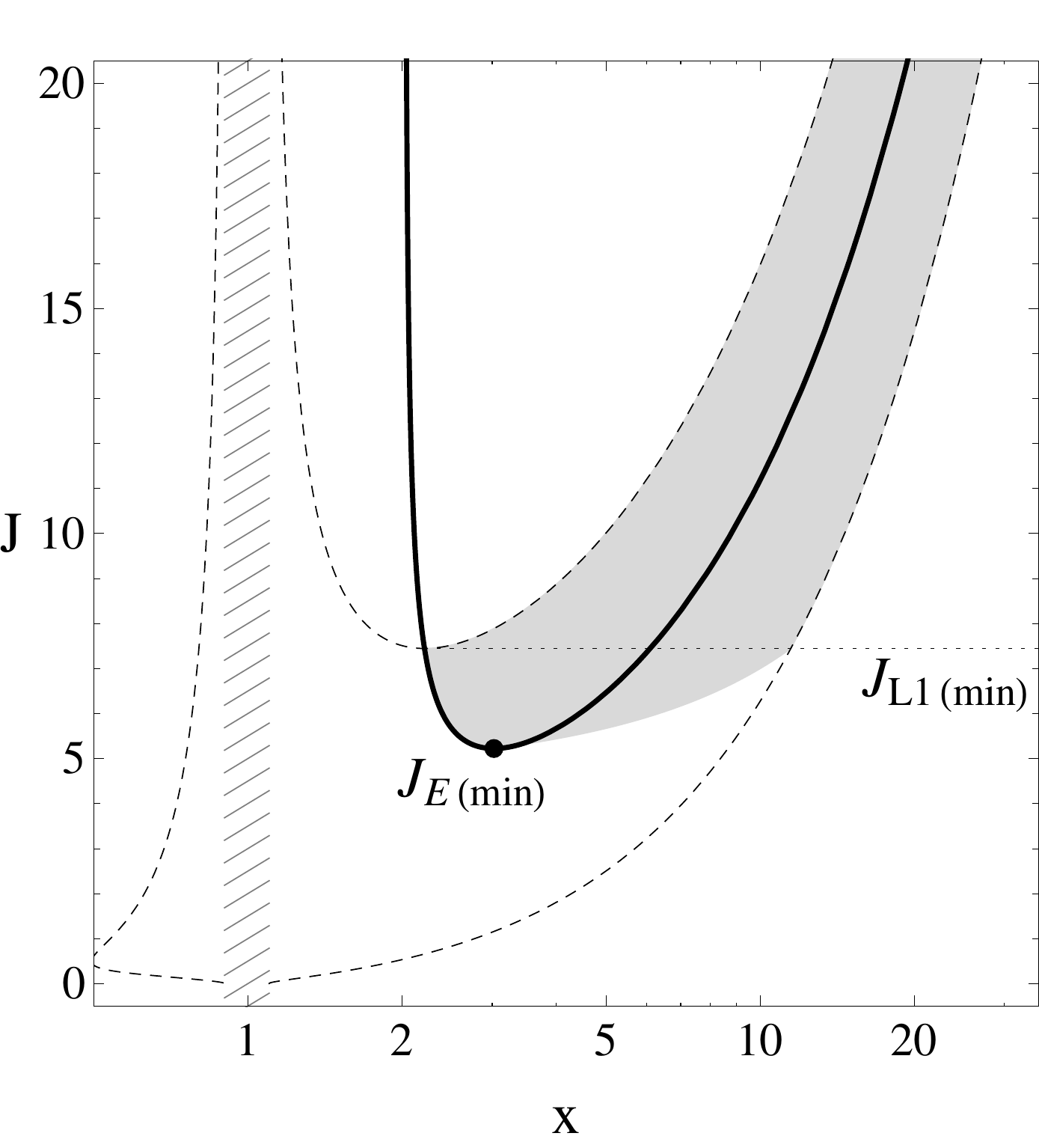}}\\
\subfigure[~ $b=1.07$ NS1]{\label{stringFIG_4c}\includegraphics[width=0.32\hsize]{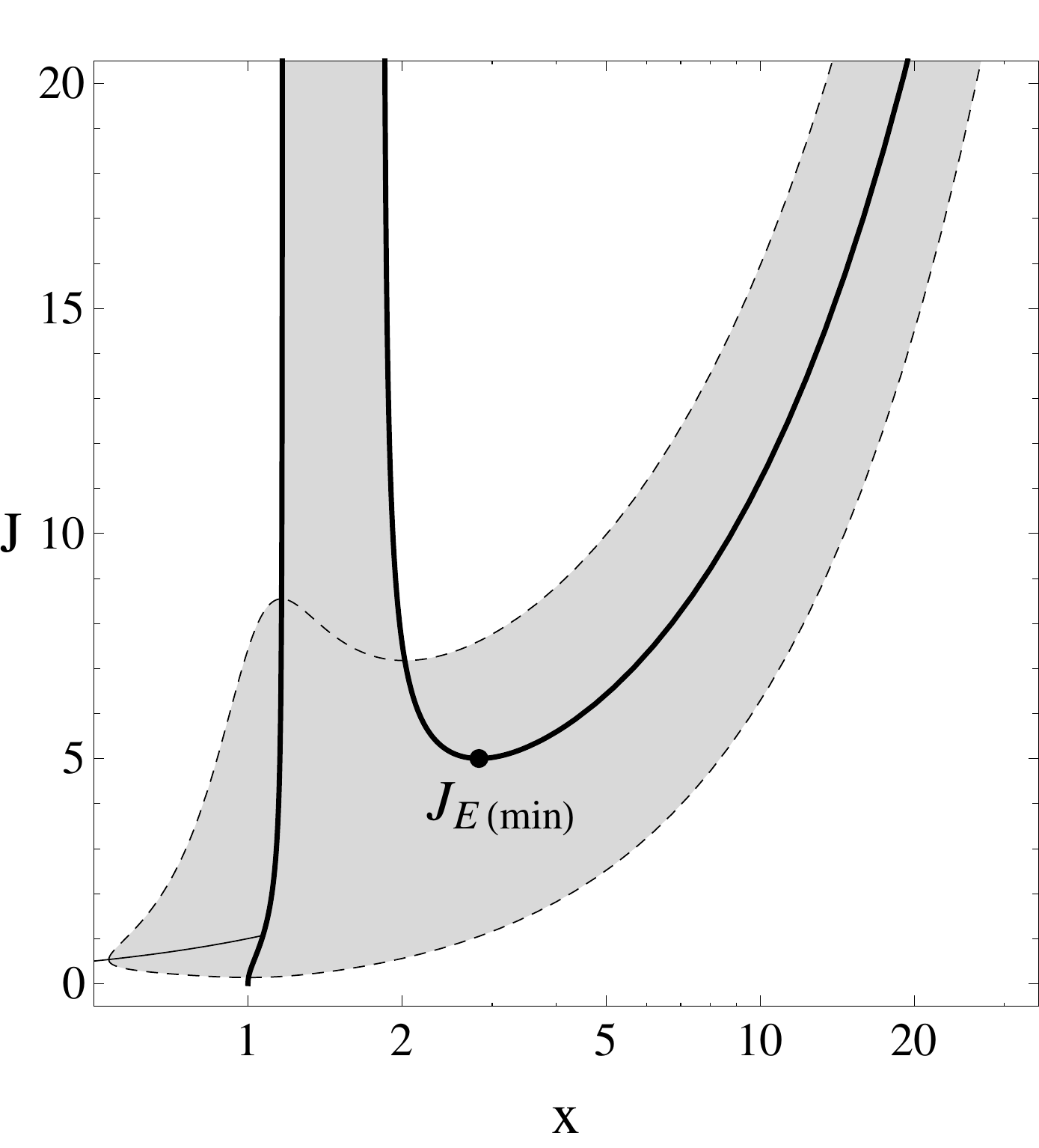}}
\subfigure[~ $b=1.14$ NS2]{\label{stringFIG_4d}\includegraphics[width=0.32\hsize]{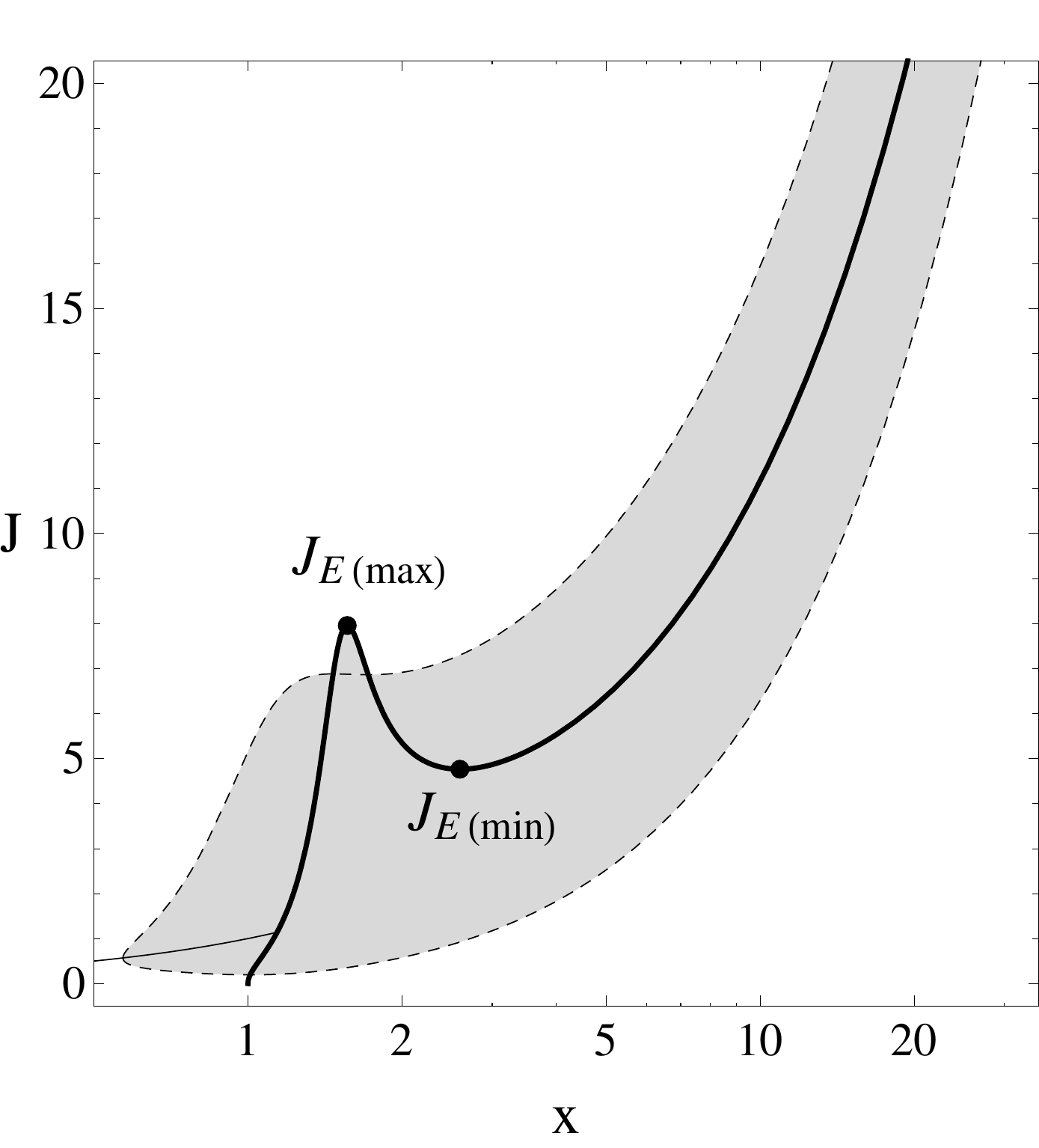}}
\subfigure[~ $b=2$ NS3]{\label{stringFIG_4e}\includegraphics[width=0.32\hsize]{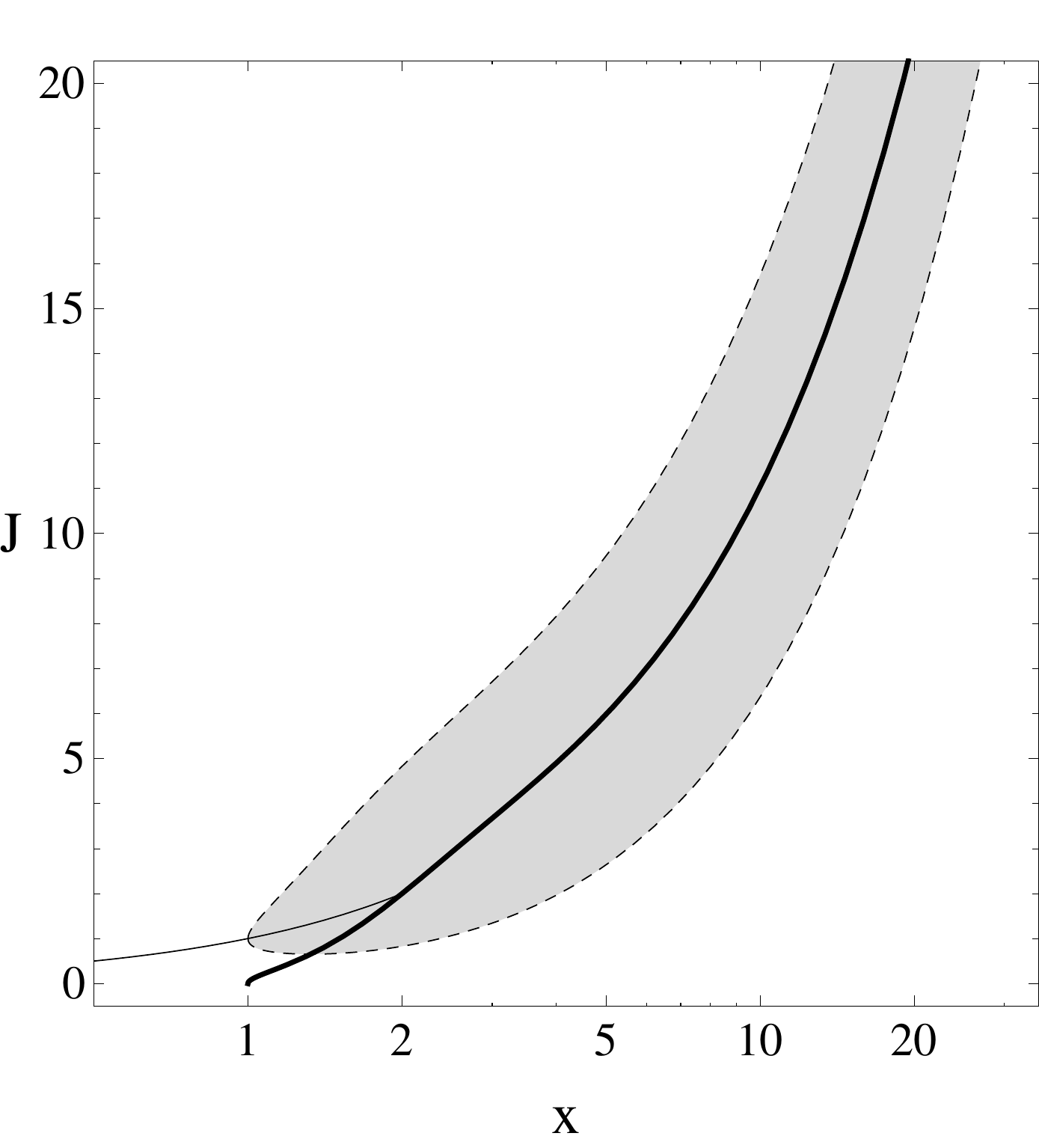}}
\vspace{-0.3cm}
\caption{\label{stringFIG_4}
Characteristic angular momentum parameter functions $\JJ(\xx;\bb)$ determining character of the boundary energy function $\EE_{\rm b}(\xx;\bb,\JJ)$, given for five different types of the braneworld RN spacetimes. The extreme function $\JJ_{\rm E}(\xx;\bb)$ (thick solid curve) determines extrema of $\EE_{\rm b}(\xx;\bb,\JJ)$. The "lake" functions $\JJ_{\rm L1}(\xx), \JJ_{\rm L2}(\xx)$ (dashed lines) determine trapped states. Regions where ``lakes'' can exist are shaded, dynamical regions between the horizons are hatched. The light full lines in the NS1-3 cases correspond to the off-equatorial minima.
}
\end{figure}

It is clear immediately that we can distinguish five qualitatively different types of the behavior of the extreme function $\JJ_{\rm E}^2(x;\bb)$ according to the value of the tidal charge of the braneworld RN spacetime; consequently, we can distinguish five types of the behavior of the energy boundary function. The function 
$\JJ_{\rm E}(x,\bb)$ is illustrated in Fig. \ref{stringFIG_3} for appropriately taken values of the tidal charge parameter $\bb$ representing the five qualitatively different types of the braneworld RN spacetimes. The extrema of the energy boundary function in the $x$-direction are given by the relations
\bea
   \EE_{\rm b(min)}(\JJ;\bb) &=& \EE_{\rm b}(\xx = x_{\rm E(min)}; \bb, \JJ), \nonumber\\  
   \EE_{\rm b(max)}(\JJ;\bb) &=& \EE_{\rm b}(\xx = x_{\rm E(max)}; \bb, \JJ). \label{Eextr}
\eea
For the five types of the RN spacetimes, we give the extrema of the energy boundary function, $\EE_{\rm b(min)}(\JJ;\bb)$ and $\EE_{\rm b(max)}(\JJ;\bb)$ in Fig. \ref{stringFIG_4}, with the tidal charges taken identically as in Fig. \ref{stringFIG_3}.

In the discussion of the properties of the string loop motion we can restrict attention to the loops with $\JJ > 0$ due to the spherical symmetry of the braneworld RN spacetimes. In the black hole RN spacetimes we focus attention to the regions outside the black hole outer horizon, while in the naked singularity RN spacetimes we consider all the region above the physical singularity, $\xx > 0, \yy > 0$. The character of the boundary energy function $\EE_{\rm b}(\xx, \yy; \bb, \JJ)$ is quite different in the $x$-direction, as compared to those in the $y$-direction, as the oscillatory character of the string loop motion is determined by the $x$-dependent part of $\EE_{\rm b}(\xx, \yy; \bb, \JJ)$. 

\begin{figure}
\subfigure[~ $y=0$]{\label{stringFIG_5a} \includegraphics[width=0.3\hsize]{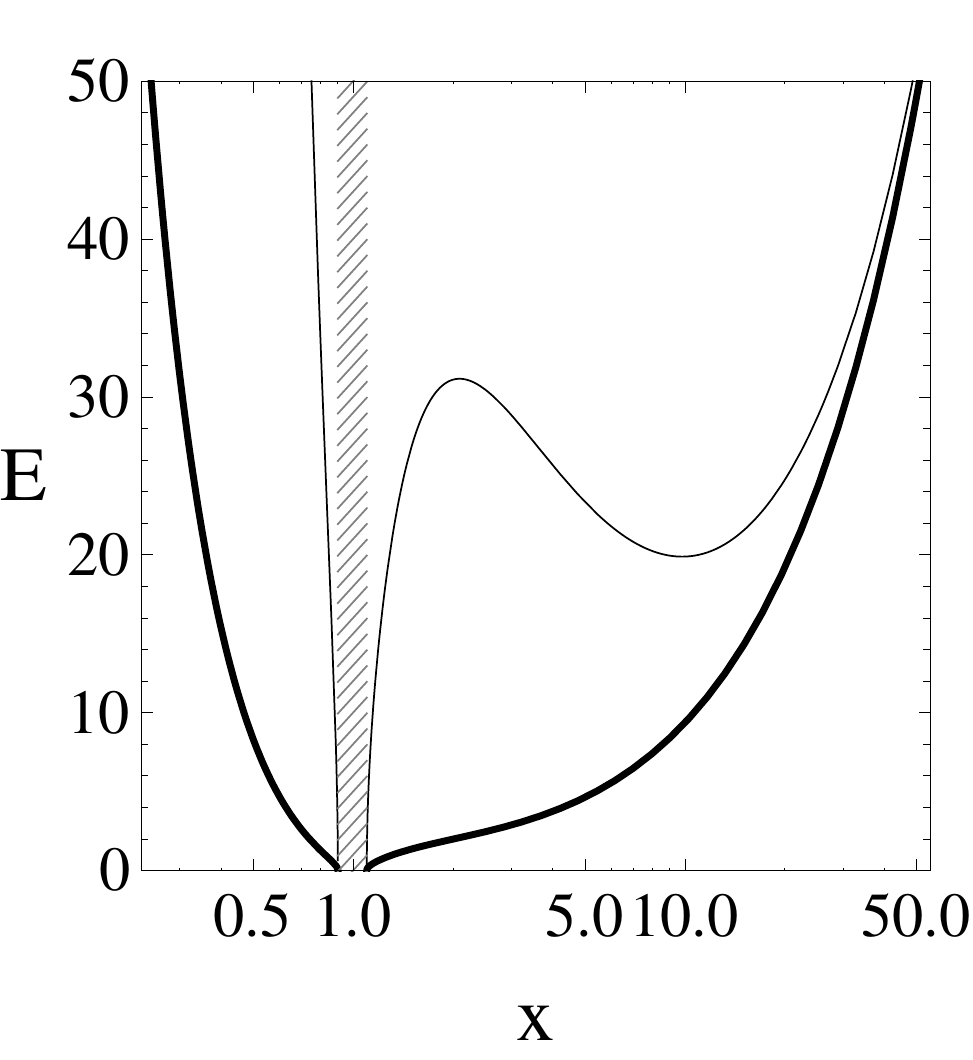}}
\subfigure[~ $x=1$]{\label{stringFIG_5b} \includegraphics[width=0.3\hsize]{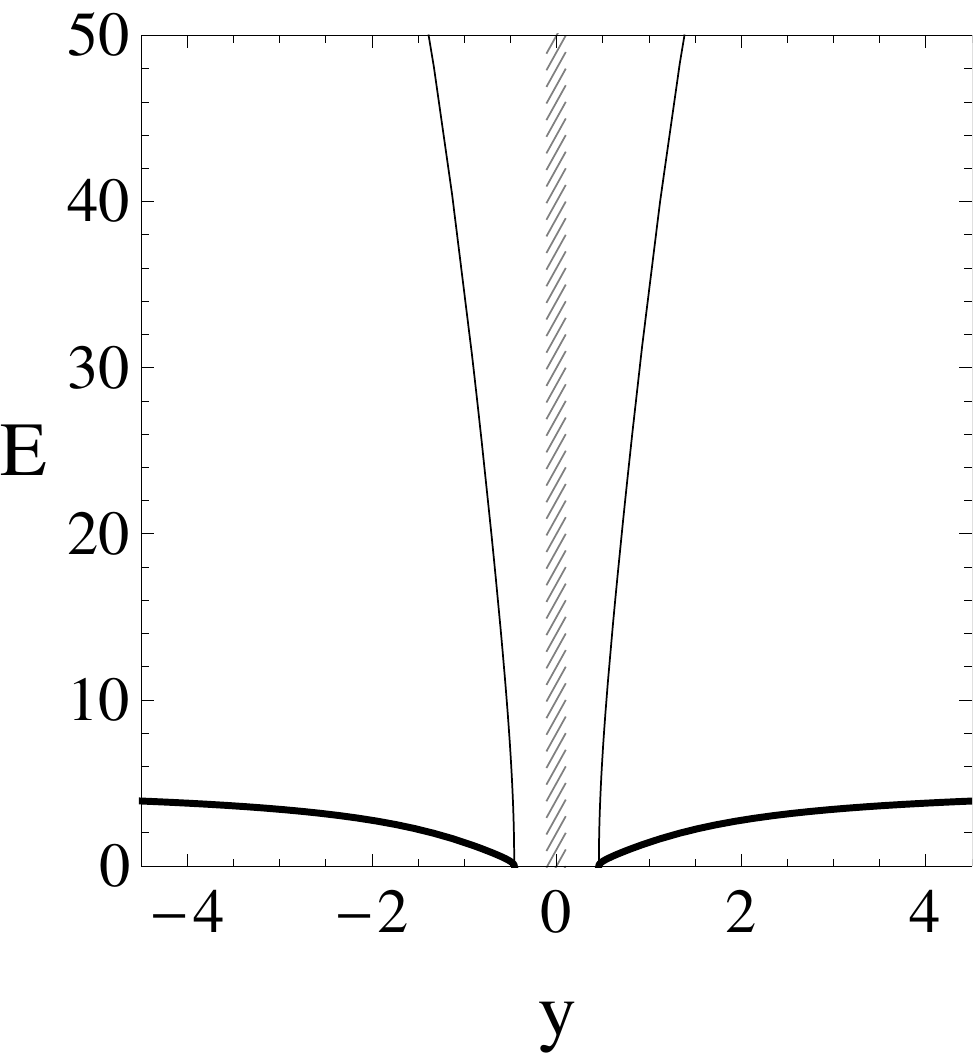}}
\vspace{-0.3cm}
\caption{\label{stringFIG_5}
Energy boundary function in braneworld RN black hole spacetime with $\bb=0.99$ represented by sections constructed for typical values of the angular momentum parameter $\JJ=11$ (thin) and $\JJ=2$ (thick). The typical sections are constructed for $y=0$, case (a), and $x=1$, case (b). For black hole spacetimes with $\bb<0$, the $\xx$-sections do not contain the inner part located under the event horizon, but remain of the same character above the event horizon.}
\subfigure[~ $y=0$]{\label{stringFIG_6a} \includegraphics[width=0.3\hsize]{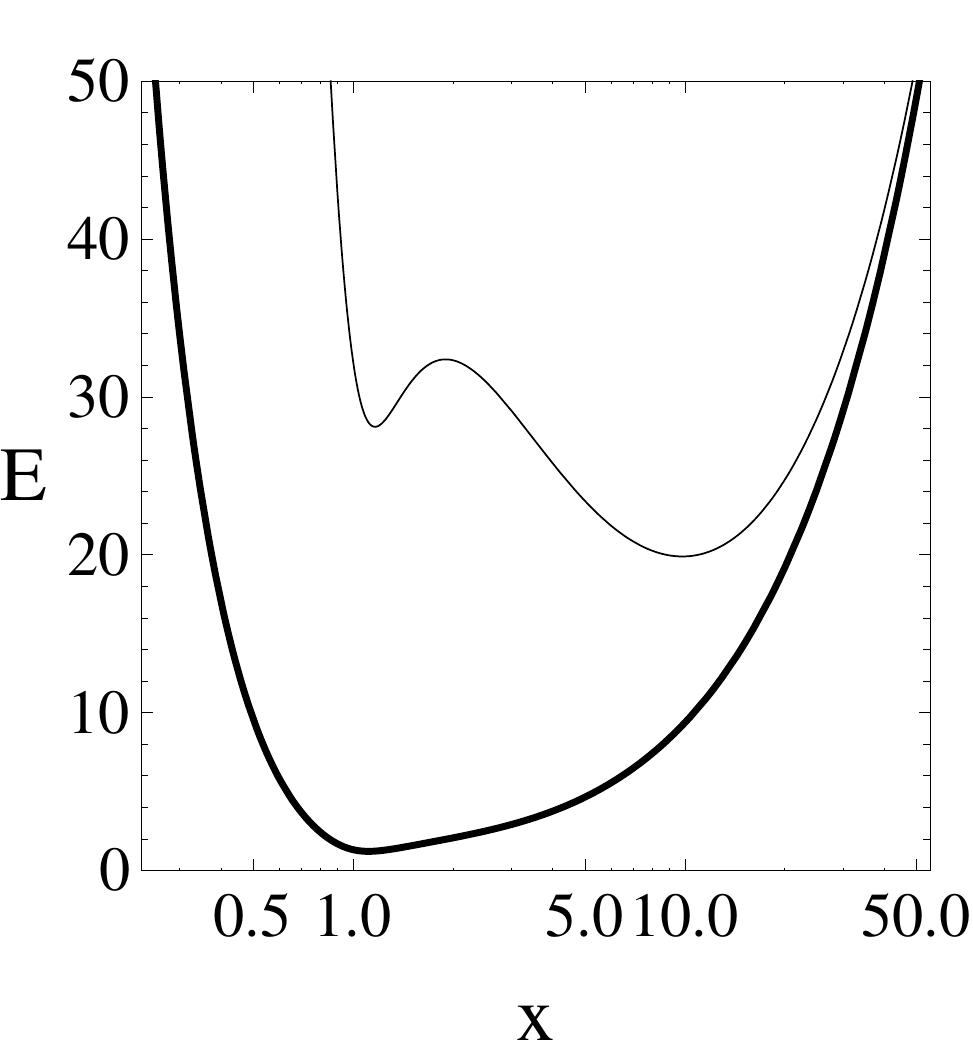}}
\subfigure[~ $x=1$]{\label{stringFIG_6b} \includegraphics[width=0.3\hsize]{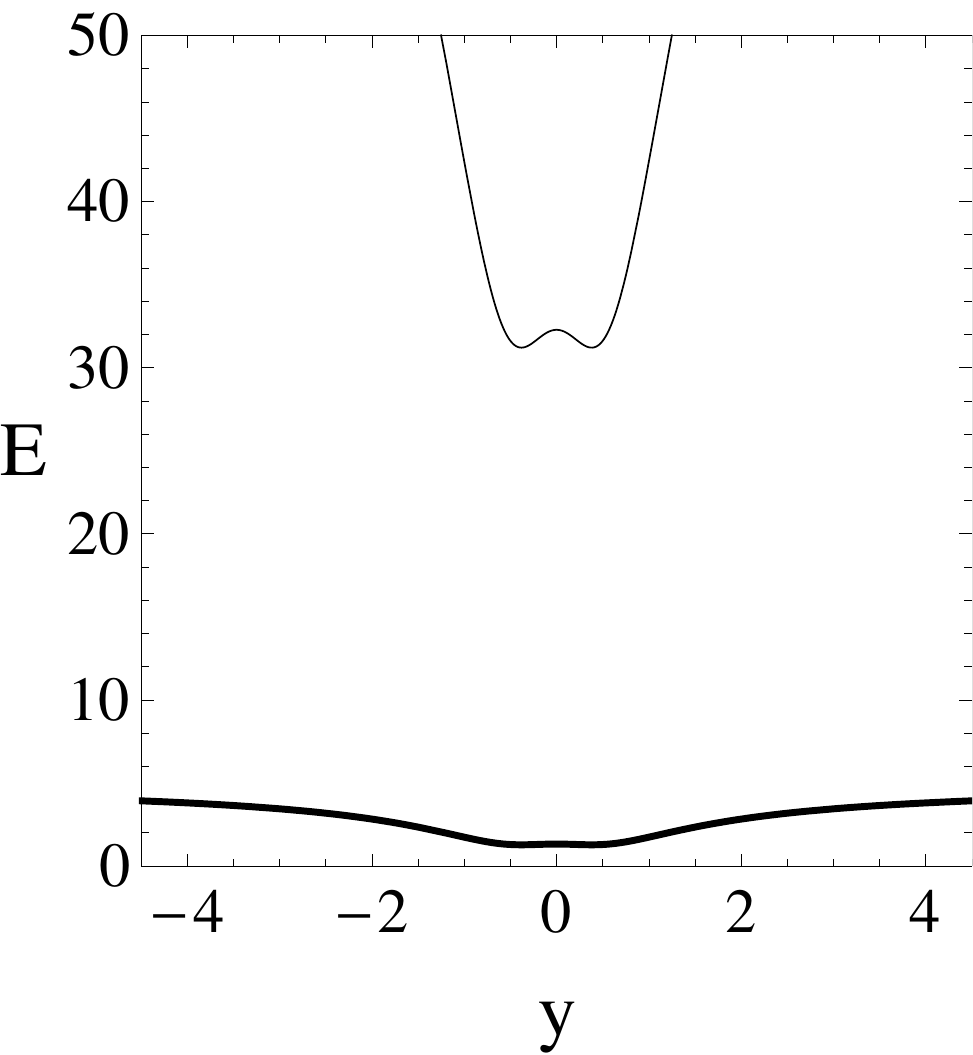}}
\vspace{-0.3cm}
\caption{\label{stringFIG_6}
Energy boundary function in braneworld RN naked singularity spacetime with $\bb=1.07$ represented by sections constructed for typical values of the angular momentum parameter $\JJ=11$ (thin) and $\JJ=2$ (thick). The sections are constructed for $y=0$, case (a), and $x=1$, case (b). In the NS3 spacetimes, only sections with one local minimum are allowed.}
\end{figure}

In the black hole RN spacetimes with both negative and positive tidal charges, the energy boundary function $\EE_{\rm b}(\xx, \yy; \bb, \JJ)$ and the equatorial extreme function $\JJ_{\rm E}^2(x;\bb)$ have the same character above the outer horizon independently on the value of $\bb<1$. For $\JJ > \JJ_{\rm E(min)}$, the energy boundary function has a local minimum and a local maximum with the extremal energy magnitudes given as a function of the string parameter $\JJ = \JJ_E$ by the relations (\ref{Eextr}), being  illustrated in Fig. \ref{stringFIG_4}. Typical $x$-profiles of the energy boundary function, $\EE_{\rm b}(\xx; \bb, \JJ)$, are illustrated in Fig. \ref{stringFIG_5a}. The oscillatory motion of the string loop in the $x$-direction is allowed for $\JJ > \JJ_{\rm E(min)}$ and energy satisfying the condition $\EE_{\rm b(min)}(\bb; \JJ) < \EE < \EE_{\rm b(max)}(\bb; \JJ)$. For $\EE < \EE_{\rm b(min)}(\bb; \JJ), \EE > \EE_{\rm b(max)}(\bb; \JJ)$ or for $\JJ < \JJ_{\rm E(min)}$, the string loop has to be captured by the black hole. The $y$-profiles of the energy boundary function, illustrated in Fig. \ref{stringFIG_5b}, determine whether the string loops are trapped in the field of the braneworld RN black hole or escape to infinity. The string loops can escape to infinity in the $y$-direction, if their energy is bigger than the minimal string energy in the flat spacetime \cite{Kol-Stu:2010:PHYSR4:} 
\beq
               \EE > \EE_{\rm min(flat)} = 2 \JJ.
\eeq

We have to determine conditions for existence and extension of regions of trapped states of the oscillating string loops in dependence on the motion constants $\JJ$ and $\EE$. Such states, in which the string loop may not escape to the infinity neither may not be captured by the black hole, correspond to ``lakes'' determined by the energy boundary function $\EE_{\rm b}(x,y;\bb,\JJ)$ for appropriately chosen energy levels. To find the "lakes", restrictions from the $x$-profiles and $y$-profiles of the boundary energy function have to be properly combined, in dependence on the string (angular momentum) parameter $\JJ$.
If the string loop motion is trapped, the condition  
\beq
           \EE_{\rm b}(\xx, \yy; \bb, \JJ) \leq 2 \JJ  \label{eq_Econ}
\eeq
has to be satisfied and we can deduce that the trapped states of the oscillating string loops could exist if 
\beq
       \JJ_{\rm L1}(\xx, \yy; \bb) > \JJ > \JJ_{\rm L2}(\xx, \yy; \bb) \label{eqTrap}
\eeq
where the so called ``lake'' angular momentum functions $\JJ_{\rm L1}(\xx, \yy; \bb)$ and $\JJ_{\rm L2}(\xx, \yy; \bb)$ are the solutions of the equality condition in (\ref{eq_Econ}) \cite{Kol-Stu:2010:PHYSR4:}. In the equatorial plane ($\yy=0$), the trapped regions are the most extended and the ``lake'' angular momentum functions take the simple form 
\beq
 \JJ_{\rm L1}(\xx;\bb) = \frac{\xx \left(\xx+\sqrt{2 \xx-\tid}\right)}{\sqrt{\xx^2-2 \xx+\tid}}, \quad \JJ_{\rm L2}(\xx;\bb) = \frac{\xx \left(\xx-\sqrt{2 \xx-\tid}\right)}{\sqrt{\xx^2-2 \xx+\tid}}. \label{eq_Lake}
\eeq
The local extrema of the functions $\JJ_{\rm L1}(\xx;\bb), \JJ_{\rm L2}(\xx;\bb)$ are given by the equation
\beq
 b^3+3 b^2 \, (r-2) r + b \, r^2 \left(r^2-10 r+12\right) + r^3 \left(-2 r^2+9 r-8\right) =  0 \label{exJ12}
\eeq
determining implicitly the function $\bb_{\rm extr(L)}(\xx)$ that is illustrated in Fig. \ref{stringFIG_1} and gives the local minima of the "lake" functions $\JJ_{\rm L1(min)}(\bb)$ that are illustrated in Fig. \ref{stringFIG_2}. The functions $\JJ_{\rm L1}(\xx;\bb)$ and $\JJ_{\rm L1}(\xx;\bb)$ are illustrated in Fig. \ref{stringFIG_3} for the typical values of the tidal charge parameter $\bb$. 

In the naked singularity RN spacetimes, we have to distinguish three qualitatively different types of the behavior of the  equatorial extreme function $\JJ_{\rm E}^2(x;\bb)$ and the energy boundary function $\EE_{\rm b}(\xx, \yy; \bb, \JJ)$ - see Fig. \ref{stringFIG_3} and Fig. \ref{stringFIG_4}; moreover, we have to consider also the off-equatorial local minima of the energy boundary function that have the same character in all three types of the naked singularity spacetimes. The energy boundary function can have in the equatorial plane two local minima and one local maximum or one local minimum for tidal charge in the range $1 < \bb < 1.202$, while only one local minimum is allowed in the spacetimes with $\bb > 1.202$. Typical $x$-profiles of the energy boundary function, $\EE_{\rm b}(\xx; \bb, \JJ)$, are illustrated for the naked singularity RN spacetimes in Fig. \ref{stringFIG_6a}. The oscillatory motion of the string loop in the $x$-direction occurs for any string parameter $\JJ > 0$ due to the asymptotic behavior of the energy boundary function at $x=0$ and the fact that a minimum of this function always exists. The condition $\EE_{\rm b(min)}(\bb; \JJ) < \EE < \EE_{\rm b(max)}(\bb; \JJ)$ can be satisfied, giving the two local minima, if $1 < \bb < 1.202$. The $y$-profiles of $\EE_{\rm b}(\xx, \yy; \bb, \JJ)$ are in the naked singularity spacetimes enriched in comparison with the black hole cases by the minima located outside the equatorial plane; they are demonstrated in Fig. \ref{stringFIG_6b}. 

As in the black hole case, the string can escape to infinity in the $y$-direction, if the string energy satisfies the condition $\EE > 2 \JJ$ reflecting the asymptotic flat spacetime limit. The states in which the string loops may not escape to infinity being trapped in the field of a braneworld RN naked singularity are governed by the condition (\ref{eqTrap}). The "lake" angular momentum functions are again given by Eq. (\ref{eq_Lake}). Their extrema are demonstrated in Fig. \ref{stringFIG_1},Fig. \ref{stringFIG_2}, Fig. \ref{stringFIG_4}. 

The structure of typical equi-energy surface sections is demonstrated in Fig. \ref{stringFIG_7} for both braneworld RN black hole and naked singularity spacetimes. We can distinguish eight qualitatively different types of the equi-energy surface sections reflecting different possibilities for the behavior of the string loops carrying an angular momentum. 


\subsection{Clasification of the braneworld RN spacetimes}

\begin{figure*}
\includegraphics[width=0.23\hsize]{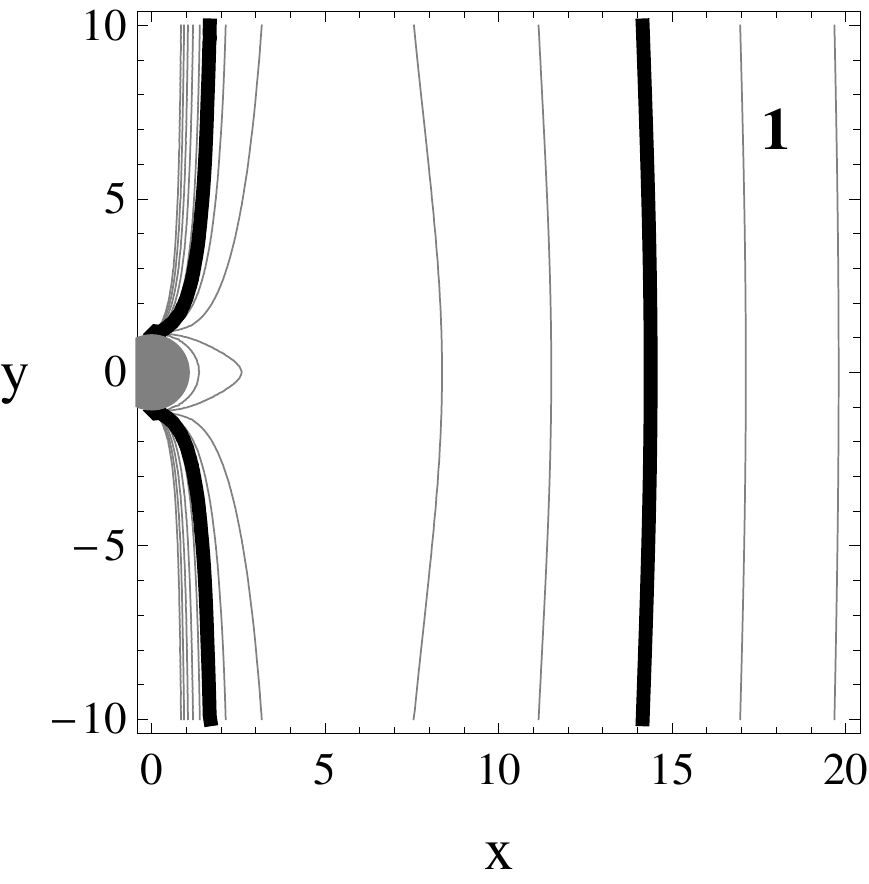}
\includegraphics[width=0.23\hsize]{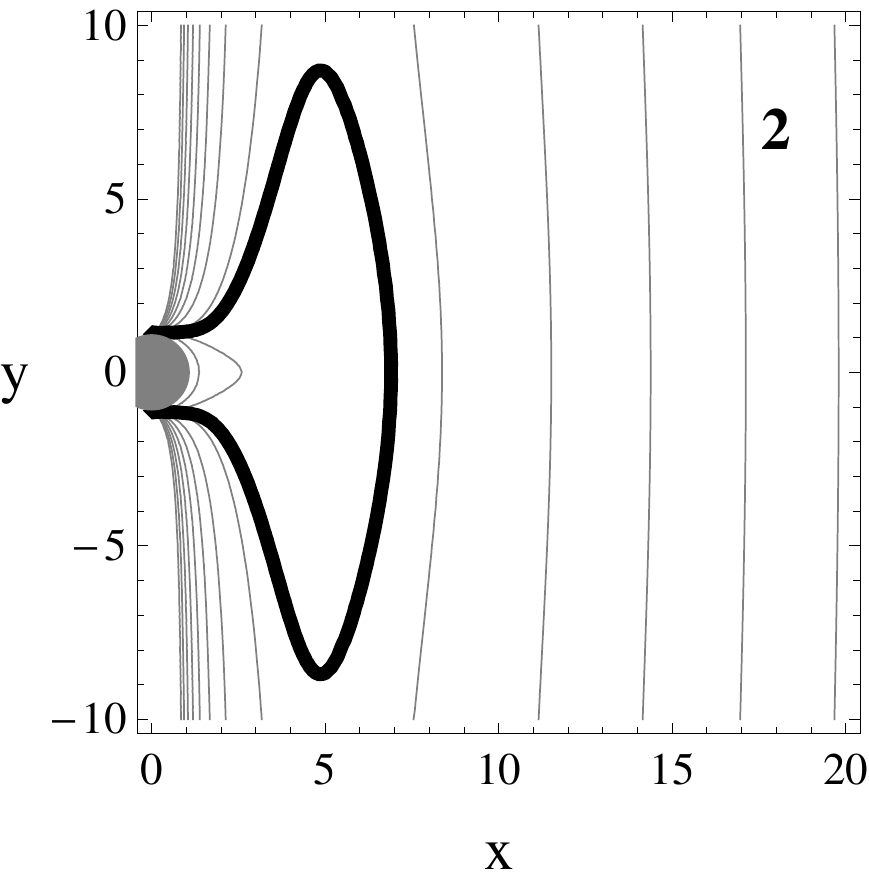}
\includegraphics[width=0.23\hsize]{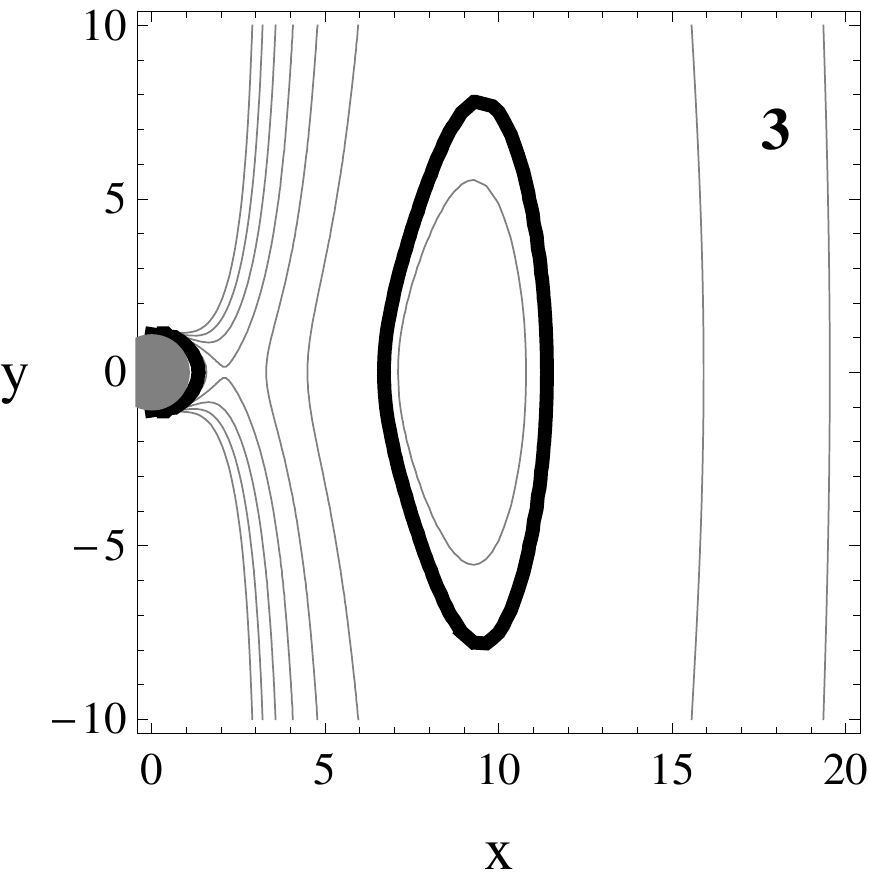}
\includegraphics[width=0.23\hsize]{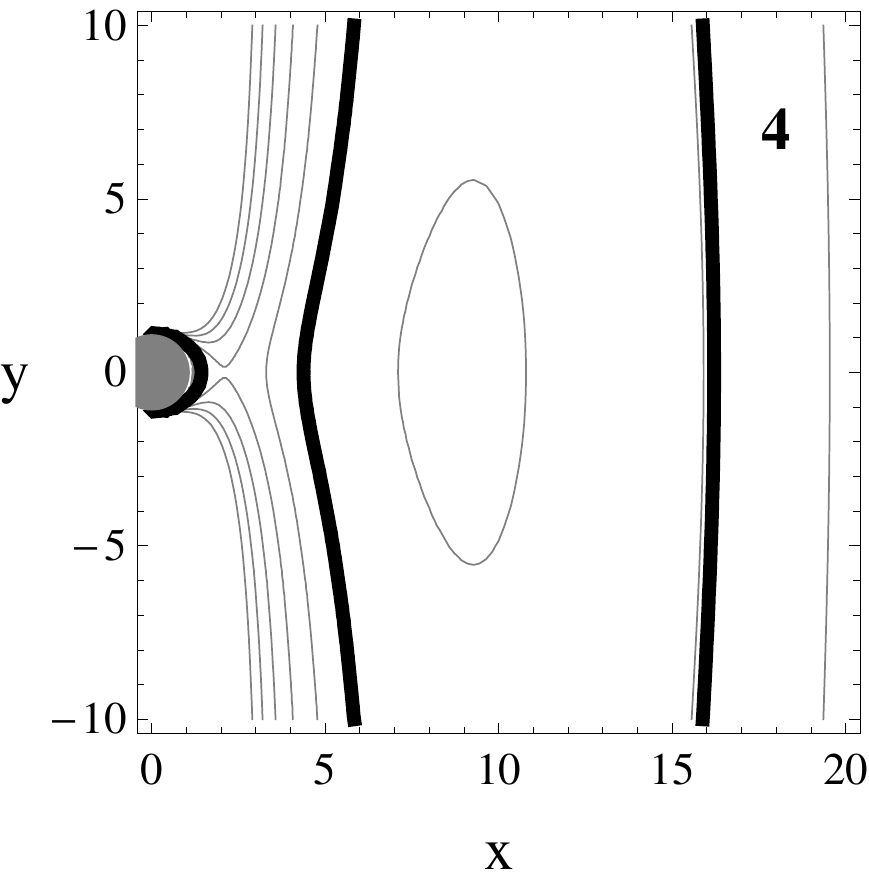}\\
\includegraphics[width=0.23\hsize]{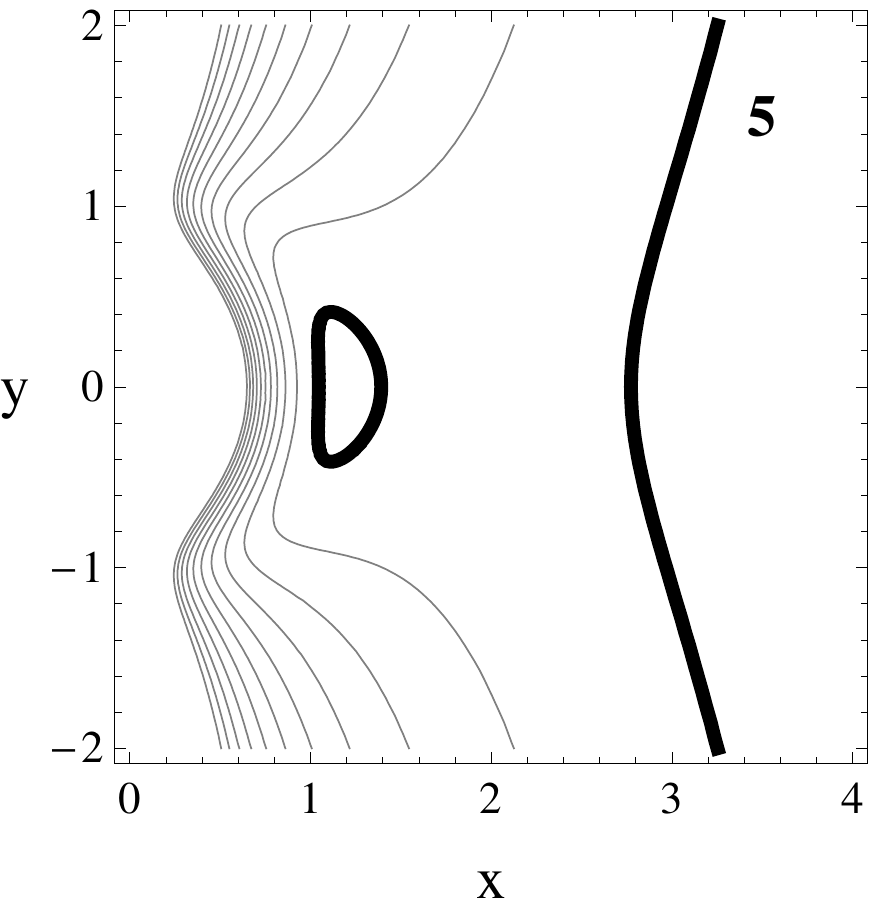}
\includegraphics[width=0.23\hsize]{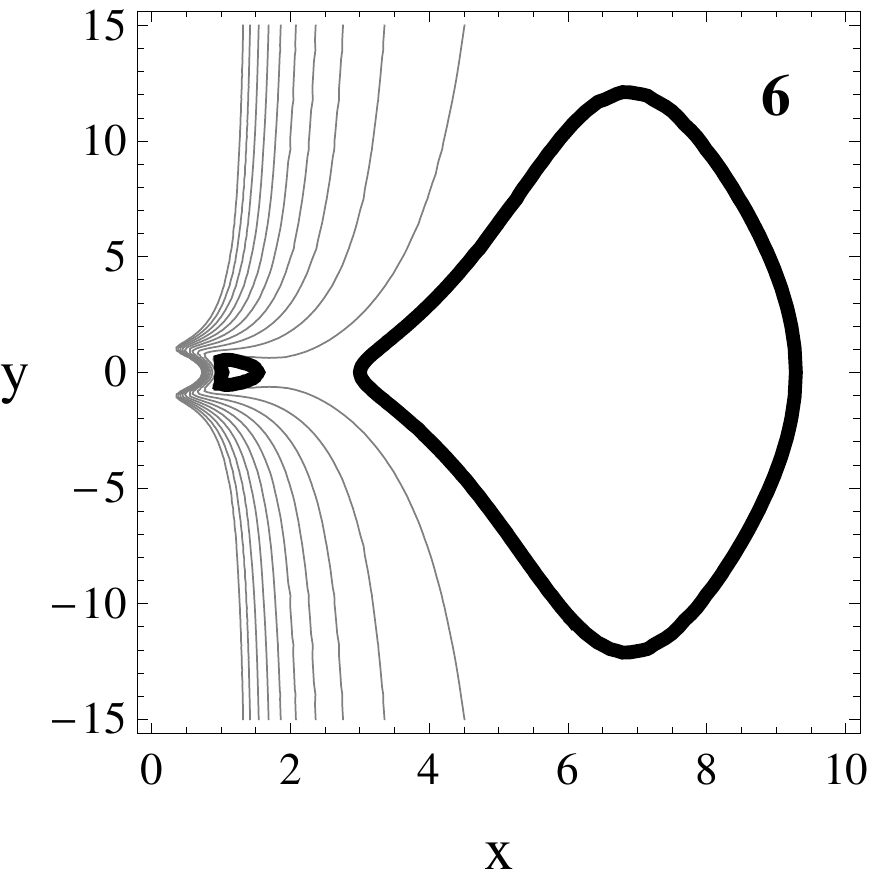}
\includegraphics[width=0.23\hsize]{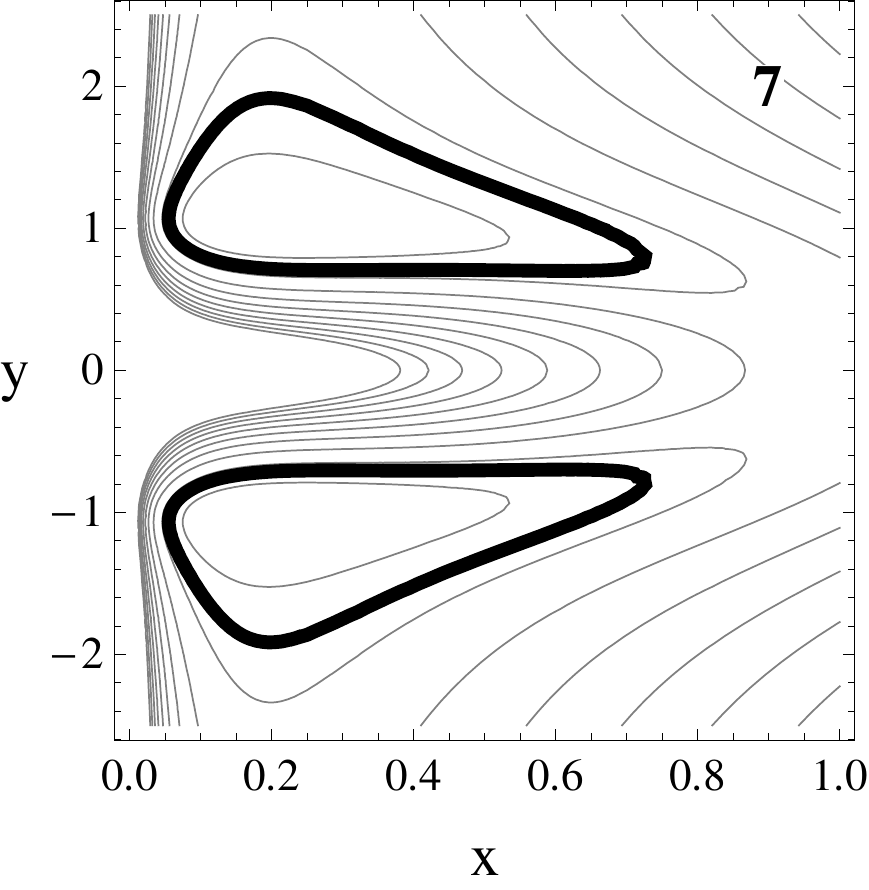}
\includegraphics[width=0.23\hsize]{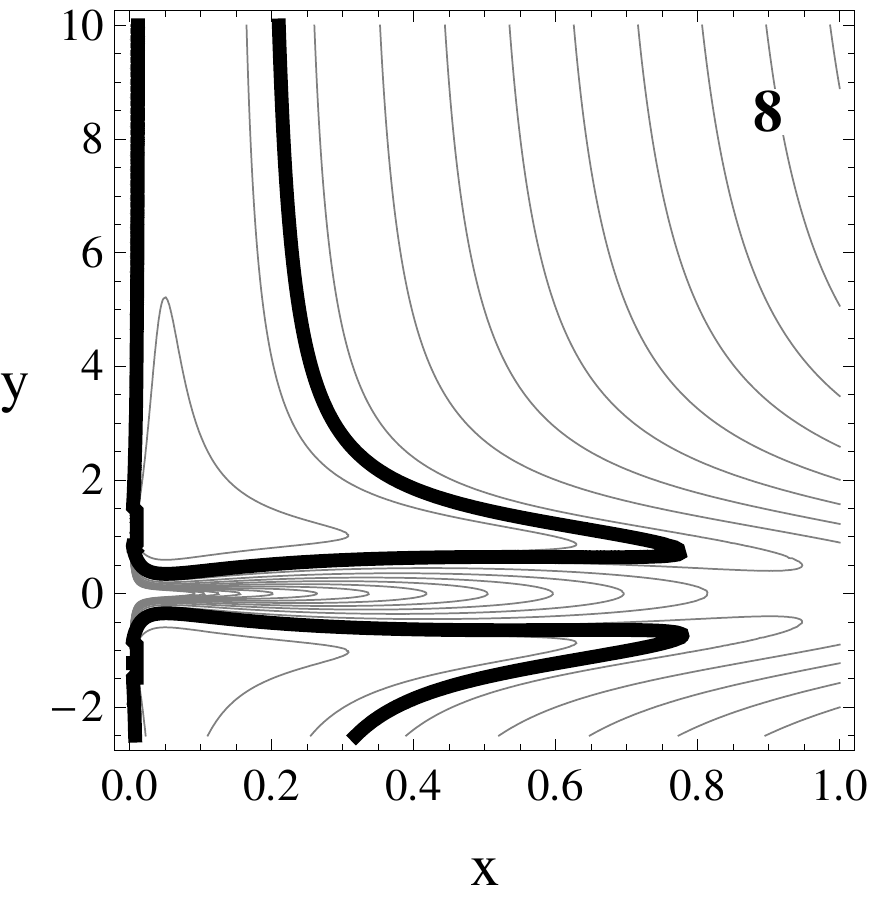}
\vspace{-0.3cm}
\caption{\label{stringFIG_7}
Eight different types of the behavior of the boundary energy function $E_{\rm b}(\xx,\yy;\bb, \JJ)$ in the braneworld RN  spacetimes represented by equi-energy surface sections. Thick contour line represents string motion boundary calculated for given string energy $\EE$ and angular momentum parameter $\JJ$, while gray lines are contours for other possible energies. Gray circle represents dynamical region below the horizon.
{\bf Case 1} ($\EE=15,\JJ=5$) (BH only) There are no inner and outer boundary and the string can be captured by the black hole or escape to infinity in the $\yy$-direction. {\bf Case 2} ($\EE=9,\JJ=5$) (BH only) The string loop cannot escape to infinity, and it must be captured by the black hole. {\bf Case 3} ($\EE=18.4,\JJ=10$) Both inner and outer boundary exist and the string is trapped in some region forming a potential ``lake'' in $\EE_{\rm}(\xx,\yy)$ around the black hole. {\bf Case 4} ($\EE=21,\JJ=10$) String cannot fall into the black hole but it can escape to infinity.
{\bf Case 5} ($\EE=30,\JJ=11$) (NS only) String can be trapped in the region around the first (inner) minimum, while it can escape to infinity from the region near the second (outer) minimum. These two regions are not connected. {\bf Case 6} ($\EE=13,\JJ=7$) (NS only) There are two disconnected regions trapping the string loops around the first or the second minimum. For the string energy high enough these two lakes can merge - corresponding to the cases 3 or 4. {\bf Case 7} ($\EE=0.2,\JJ=0.2$) (NS only) Two disconnected lakes located  off the equatorial plane. String can not escape from this regions. {\bf Case 8} ($\EE=0.2,\JJ=0.05$)  (NS only) Two disconnected regions exist. String loops can escape to infinity, but can not cross the equatorial plane.  
Cases 3 and 4 can appear in both BH and NS spacetimes.
}
\end{figure*}

We discuss in a systematic way the five qualitatively different types of the behavior of the effective potential (or, equivalently, the energy boundary function) characterizing the string loop motion in the braneworld RN spacetimes. There are two types of the black hole spacetimes and three types of the naked singularity spacetimes, distinguished by the interval of the braneworld tidal charge parameter $\bb$.
\begin{itemize}
\item {\bf BH-Negative} ($\bb<0$) \\ 
Braneworld RN black holes with only one event horizon. Only one solution of the equation $\bb=\bb_{\rm ex}(\xx)$ exists, which implies existence of only one minimum $\JJ_{\rm E(min)}$ of the extremal angular momentum function $\JJ_{\rm E}(\xx;\bb)$  - see Fig. \ref{stringFIG_1}. For $\JJ>\JJ_{\rm E(min)}$, there exist a local maximum and minimum of the energy boundary function $\EE_{\rm b}$ in the equatorial plane. For $\JJ<\JJ_{\rm E(min)}$, no local extrema exist - see Fig. \ref{stringFIG_3a}. In these spacetimes we obtain four different types of the string loop motion corresponding the captured, trapped, scattered and rescattered motion that are reflected by the four different types of the equi-energy surface sections - see figures 1-4 in Fig \ref{stringFIG_7}, and Fig. \ref{stringFIG_3a}. The situation is of the same character as in the Schwarzschild spacetime \cite{Kol-Stu:2010:PHYSR4:}.
\item {\bf BH-Positive} ($0<\bb \leq 1$) \\ 
Braneworld \RN{} black holes with two event horizons. Above the astrophysically relevant outer horizon, the character of the string loop motion is the same as in the BH-Negative spacetimes, demonstrating the four types of the motion - see figures 1-4 from Fig \ref{stringFIG_7}; there are quantitative differences only, shifting the motion to the deeper gravitational field in vicinity of the outer horizon - see Fig. \ref{stringFIG_3b}. On the other hand, under the inner event horizon, there are no local extrema of the energy boundary function. We shall not consider motion under the inner horizon in the following. 
\item {\bf NS1} ($1<\bb<9/8$) \\ 
\RN{} naked singularities allowing for two solutions of the equation $\bb=\bb_{\rm ex}(x)$ and two solutions of the equation $\bb=\bb_{\rm div}(x)$ (Fig. \ref{stringFIG_1}), which implies existence of one local minimum $\JJ_{\rm E(min)}$ of the extremal angular momentum function $\JJ_{\rm E}(\xx;\bb)$ function in addition to an internal branch of this function covering the whole range of $\JJ>0$ (Fig. \ref{stringFIG_3}). For $\JJ>\JJ_{\rm E(min)}$ there exist two local minima and one local maximum of the energy boundary function $\EE_{\rm b}(\xx; \bb, \JJ)$; for $\JJ<\JJ_{\rm E(min)}$, only the inner local minimum remains. Then Fig. \ref{stringFIG_4c} and properties of the off-equatorial local minima of the energy boundary function imply that in this type of the spacetimes we can recognize six different types of the equi-energy surface sections of the boundary energy function denoted as cases ($3 - 8$) in Fig \ref{stringFIG_7}. The cases ($3 - 4$) are identical to the cases relevant in the black hole spacetimes and correspond to the trapped and escaping string loop motion. The cases ($5 - 8$) are typical for the naked singularity spacetimes. In the case ($5$), there is an inner region of trapped motion and an outer region of the escaping motion allowing both for scattering and rescattering of the string loops. In the case ($6$), there are two regions of the trapped motion. Finally, we have two cases related to the off-equatorial minima of the motion. In the case ($7$), two trapped regions symmetrically distributed above and under the equatorial plane exist. In the case ($8$), two regions of the escaping motion are symmetrically distributed above and under the equatorial plane, allowing only for rescattering of the string loops. Crossing of the equatorial plane is forbidden in the cases ($7$) and ($8$).  
\item {\bf NS2} ($9/8<\bb<\bb_{\rm ex}(\xx_{\rm max})$) \\ 
Two solutions of the equation $\bb=\bb_{\rm ex}(x)$ exist and no solution of $\bb=\bb_{\rm div}(x)$ (Fig. \ref{stringFIG_1}), implying existence of one local minimum $\JJ_{\rm E(min)}$ and one local maximum $\JJ_{\rm E(max)}$ of the extremal angular momentum function $\JJ_{\rm E}(\xx,\bb)$ (Fig. \ref{stringFIG_3d}). For $\JJ>\JJ_{\rm E(max)}$ and $\JJ_{\rm E(min)}>\JJ$ only one local minimum of the energy boundary function $\EE_{\rm b}(\xx;\bb,\JJ)$ exists, while for $\JJ_{\rm E(max)}>\JJ>\JJ_{\rm E(min)}$ there are two local minima and one local maximum of the energy boundary function. In NS2 spacetimes we recognize the same six different types of the equi-energy surface sections as in the case of NS1 spacetimes. 
\item {\bf NS3} ($\bb>\bb_{\rm ex}(\xx_{\rm max})$) \\ 
No solution of the equations $\bb=\bb_{\rm ex}(x)$ and $\bb=\bb_{\rm div}(x)$ exist, implying existence of no local extremum of the extremal angular momentum function $\JJ_{\rm E}(\xx,\bb)$ - see Fig. \ref{stringFIG_1} and Fig. \ref{stringFIG_3e}. There is only one local minimum of the energy boundary function $\EE_{\rm b}(\xx;\bb,\JJ)$ - see Fig. \ref{stringFIG_4e}. In the NS3 spacetimes we recognize four different types of the equi-energy surface sections represented by the cases ($3,4,7,8$) in Fig \ref{stringFIG_7}. 
\end{itemize}

We can summarize that in the case of black holes ($\bb \leq 1$) the string loops can collapse to the black hole reaching the event horizont and the physical singularity at $r=0$ for appropriately chosen angular momentum parameter and energy. There are four qualitatively different cases of the string loop motion in the braneworld RN black hole spacetimes reflecting the capturing, trapping or escaping of the oscillating string loops. In the case of naked singularity spacetimes ($\bb>1$), the string can not reach the spacetime singularity at $r=0$ since   
\beq
 \lim_{x \rightarrow 0} \EE_{\rm b}(\xx,\yy) = \infty 
\eeq
for all $\yy$ - there is always a repulsive barrier arising due to the string loop angular momentum parameter $\JJ$.
For string loops it is impossible to collapse into the naked singularity and they have to be in trapped states near the naked singularity or be able to escape to infinity along the $\yy$-direction. There are six qualitatively different types of the string loop motion in the braneworld RN naked singularity spacetimes reflecting trapping or escaping of string loops. 

\section{String loop acceleration and asymptotical ejection speed}

>From the astrophysical point of view, one of the most relevant applications of the axisymmetric string loop motion is the possibility of strong acceleration of the linear string motion due to the transmutation process in the strong gravity of extremely compact objects that occurs due to the chaotic character of the string loop motion and could well mimic acceleration of relativistic jets in Active Galactic Nuclei (AGN) and microquasars \cite{Jac-Sot:2009:PHYSR4:,Stu-Kol:2012:PHYSR4:}. Since the braneworld RN spacetimes are asymptotically flat, we first discuss the string loop motion in the flat spacetime enabling clear definition of the acceleration process. The energy of the string loop (\ref{StringEnergy}) in the flat spacetime expressed in the Cartesian coordinates reads
\beq
E^2 = \dot{y}^2 + \dot{x}^2 + \left( \frac{J^2}{x} + x \right)^2  =  E^2_{\mathrm y} + E^2_{\mathrm x}, \label{E2flat}
\eeq
where dot denotes derivative with respect to the affine parameter $\af$. The energies related to the $x$- and $y$-directions are given by the relations
\beq
  E^2_{\mathrm y} = \dot{y}^2, \quad E^2_{\mathrm x} = \dot{x}^2 + \left( \frac{J^2}{x} + x \right)^2 = 
                                                          (x_{\rm i} + x_{\rm o})^2 = E^2_{0}  \label{restenergy}
\eeq
where $x_i$ ($x_o$) represent the inner (outer) limit of the oscillatory motion. 
The energy $E_0$ representing the internal energy of the string loop is minimal when the inner and the outer radii coincide, leading to the relation 
\beq
 E_{\rm 0(min)} = 2 J \label{E0min}
\eeq 
that determines the minimal energy necessary for escaping of the string loop to infinity in the spacetimes related to black holes or naked singularities. Clearly, $E_{\rm x}=E_{0}$ and $E_{\rm y}$ are constants of the string loop motion and no transmutation between these energy modes is possible in the flat spacetime. However, in vicinity of black holes, the kinetic energy of the oscillating string can be transmitted into the kinetic energy of the translational linear motion (or vice versa) due to the chaotic character of the string loop motion.

\begin{figure*}
\includegraphics[width=\hsize]{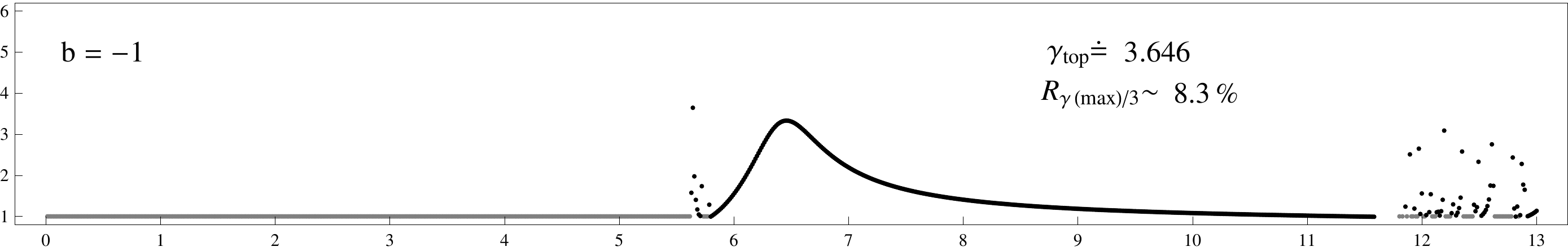}
\includegraphics[width=\hsize]{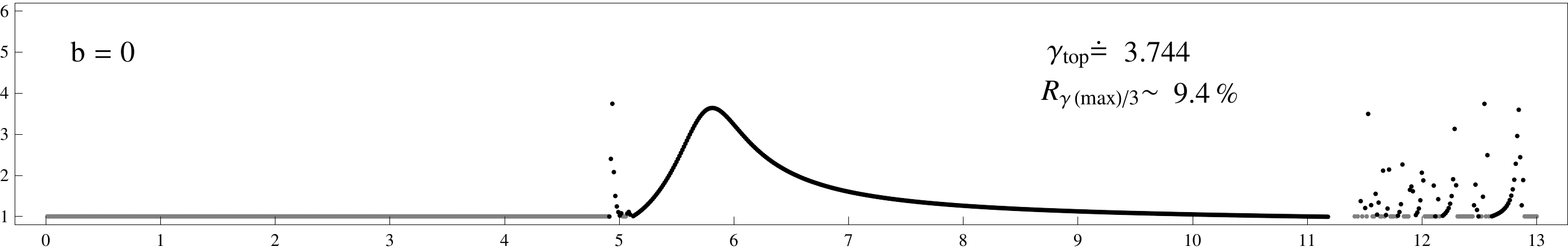}
\includegraphics[width=\hsize]{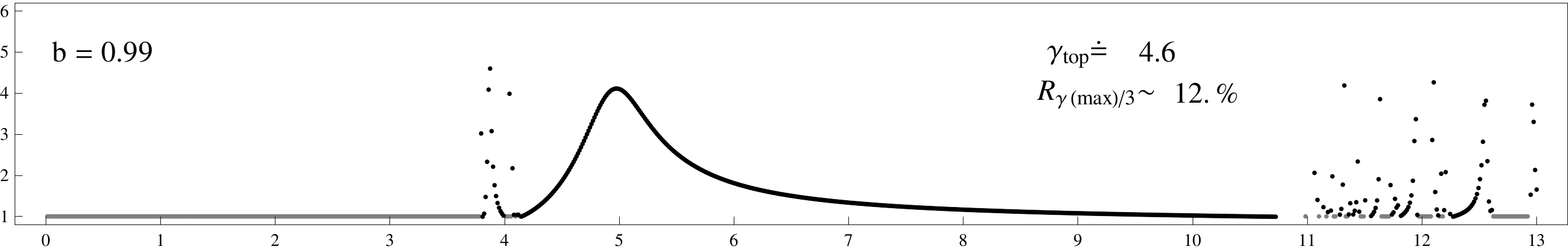}
\includegraphics[width=\hsize]{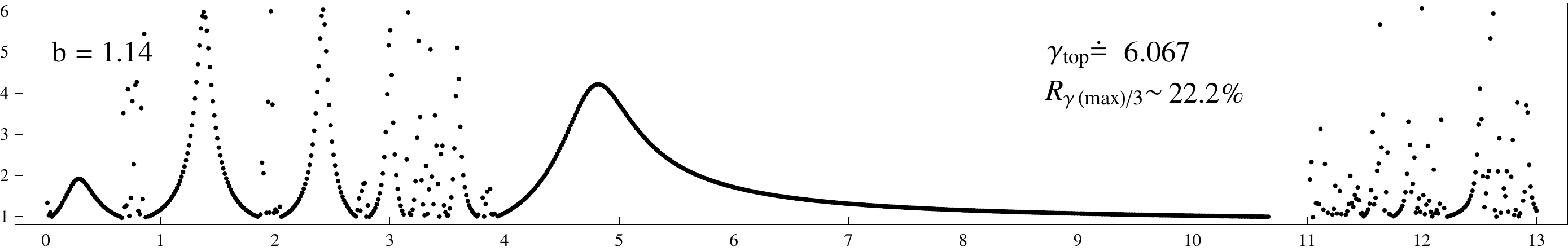}
\includegraphics[width=\hsize]{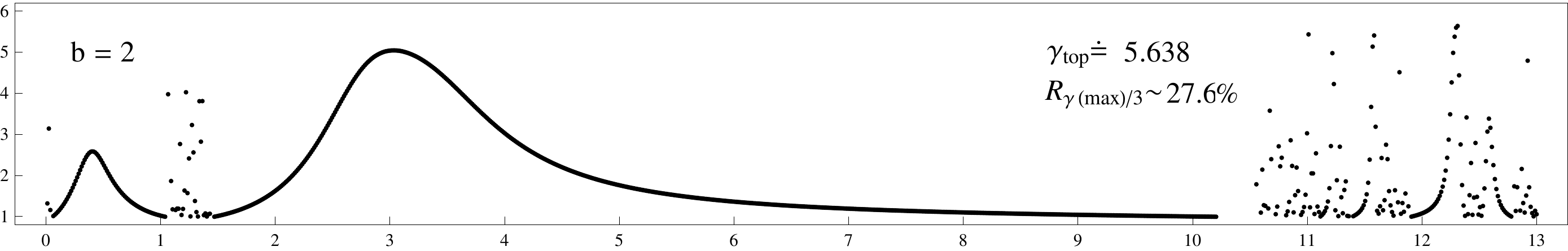}
\vspace{-0.3cm}
\caption{\label{stringFIG_8}
The asymptotic speed of transmitted string loops for different values of parameter $\bb$. The Lorentz factor $\gamma$ (vertical axis) is calculated for strings with energy $\EE=25$ and current $\JJ=2$, starting from the rest with different initial position $\yy_0 \in (0,13) $ (horizontal axis) while $\xx_0$ is calculated from (\ref{EqEbXY}). Maximal acceleration (\ref{gmmax}) for this case gives us the limiting gamma factor $\gamma_{\rm max} = 6.25$. We show the topical  gamma factor that is numerically found in the sample, $\gamma_{\rm top}$, and also the efficiency of the transmutation effect, $R_{\gamma(\rm max)/3}$, given by the relative number of accelerations when the final Lorentz factor is larger than $\gamma_{\rm max}/3$.
}
\end{figure*}

\begin{figure*}
\includegraphics[width=\hsize]{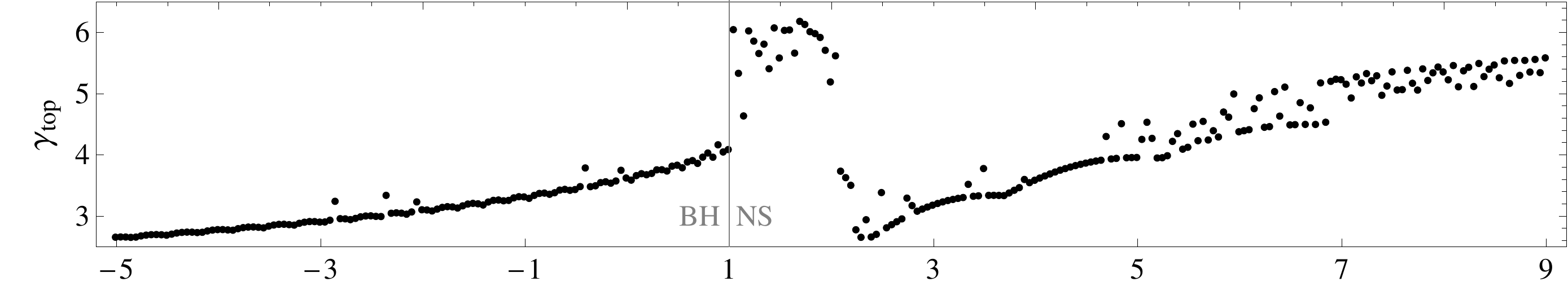}
\includegraphics[width=\hsize]{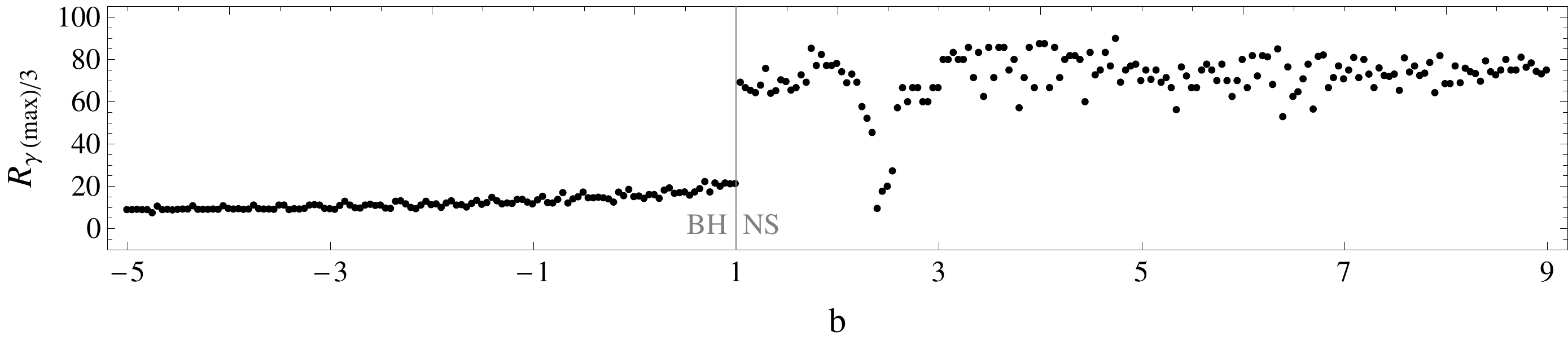}
\vspace{-0.3cm}
\caption{\label{stringFIG_9}
Topical string loop speed $\gamma_{\rm top}$ and the relative efficiency factor $R_{\gamma(\rm max)/3}$ in dependence on the tidal charge $\bb$. Notice the sharp change at BH/NS transition. String loops are assumed with energy $\EE=25$ and current $\JJ=2$, maximal acceleration is limited by $\gamma_{\rm max} = 6.25$.}
\end{figure*}

In order to get a strong acceleration in the braneworld RN spacetimes, the string loop has to pass the region of strong gravity near the black hole horizon or in vicinity of the naked singularity, where the string transmutation effect $\EE_{\rm x} \leftrightarrow \EE_{\rm y}$ can occur. In the spacetimes with $A(\rr) \neq 1$, the string loop energy (\ref{StringEnergy}) can be expressed in the Cartesian coordinates in the form
\beq
\EE^2 = A(\rr) \left( g_{xx}\dot{\xx}^2 + 2 g_{xy}\dot{\xx}\dot{\yy} + g_{yy}\dot{\yy}^2 \right) + A(\rr) \left(\frac{\JJ^2}{\xx} + \xx \right)^2, \label{Enedec1}
\eeq
where the metric coefficients of the braneworld RN spacetimes in the $\xx$ and $\yy$-coordinates are given by
\beq
 g_{xx} = \frac{\xx^2 + A \yy^2}{A (\xx^2+\yy^2)}, \quad g_{yy} = \frac{\yy^2 + A \xx^2}{A (\xx^2+\yy^2)}, \quad g_{xy} = \xx \yy \frac{1 - A}{A (\xx^2+\yy^2)}.    \label{Enedec2}
\eeq
The term $g_{xy}\dot{\xx}\dot{\yy}$ is responsible for interchange of energy between the $\EE_{\rm x}$ and $\EE_{\rm y}$ energy modes. The metric coefficient $g_{xy}$ is significant only in the strong gravity region near the black hole horizon or naked singularity, therefore, the effect of string transmutation can occur only in these strong gravity regions. 

All energy of the transitional ($\EE_{\rm y}$) energy mode can be transmitted to the oscillatory ($\EE_{\rm x}$) energy mode - oscillations of the string loop in the $\xx$-direction and the internal energy of the string will increase maximally in such a situation, while the string will stop moving in the $\yy$-direction. However, all energy of the $\EE_{\rm x}$ mode cannot be transmitted into the $\EE_y$ energy mode - there remains inconvertible internal energy of the string,  $\EE_{\rm 0(min)}=2\JJ$, being the minimal potential energy hidden in the $\EE_{\rm x}$ energy mode. It should be stressed that no rotation of the black hole (naked singularity) is necessary for acceleration of the string loop motion due to the transmutation effect, contrary to the Blandford---Znajek effect \cite{Bla-Zna:1977:MNRAS:} usually considered in modelling acceleration of jet-like motion in AGN and microquasars.

The final Lorentz factor of the transitional motion of an accelerated string loop as observed in the asymptotically flat region of the braneworld RN spacetimes is due to (\ref{restenergy}) determined by the relation \citep{Jac-Sot:2009:PHYSR4:,Stu-Kol:2012:PHYSR4:} 
\beq
 \gamma = \frac{\EE}{\EE_0} = \frac{\EE}{\xx_{\rm i} + \xx_{\rm o}}, \label{gamma}
\eeq 
where $\EE$ is the total energy of the string loop moving with the internal energy $\EE_{0}$ in the $\yy$-direction with the velocity corresponding to the Lorentz factor $\gamma$ - for details see \cite{Stu-Kol:2012:PHYSR4:}. 

\beq
 \gamma_{\rm max} = \frac{\EE}{2 \JJ} \label{gmmax} .
\eeq

We have tested the efficiency of the transmutation effect by studying the string loop acceleration and ejection speed in dependence on braneworld tidal charge spacetime parameter and the string loop parameters. First, we have tested for fixed string parameters ($\EE=25, \JJ=2$) the results of the transmutation effect in spacetimes with typically chosen values of the tidal charge parameter $\bb$, covering the five cases of the behavior of the energy boundary function in the braneworld RN spacetimes. The results are illustrated in Fig. \ref{stringFIG_8}, where we also give the extremal Lorentz factor obtained by the numerical experiments reflecting the chaotic character of the string loop motion, $\gamma_{\rm top}$, being available for the given string parameters in the given spacetime; we further present the relative ratio $R_{\gamma(\rm max)/3}$ of the numerical experiments leading to the final Lorentz factor $\gamma > \gamma_{\rm max}/3$ that is also  reflecting the efficiency of the transmutation process. The dependence of the topical Lorentz factor and the relative ratio of strongly accelerated string loops on the tidal charge of the RN spacetimes are illustrated in Fig. \ref{stringFIG_9}. 

\begin{figure}
\hspace{-1.0cm}
\subfigure[\,RN BH \, $\bb = -1$]{\label{stringFIG_10a}\includegraphics[width=0.4\hsize]{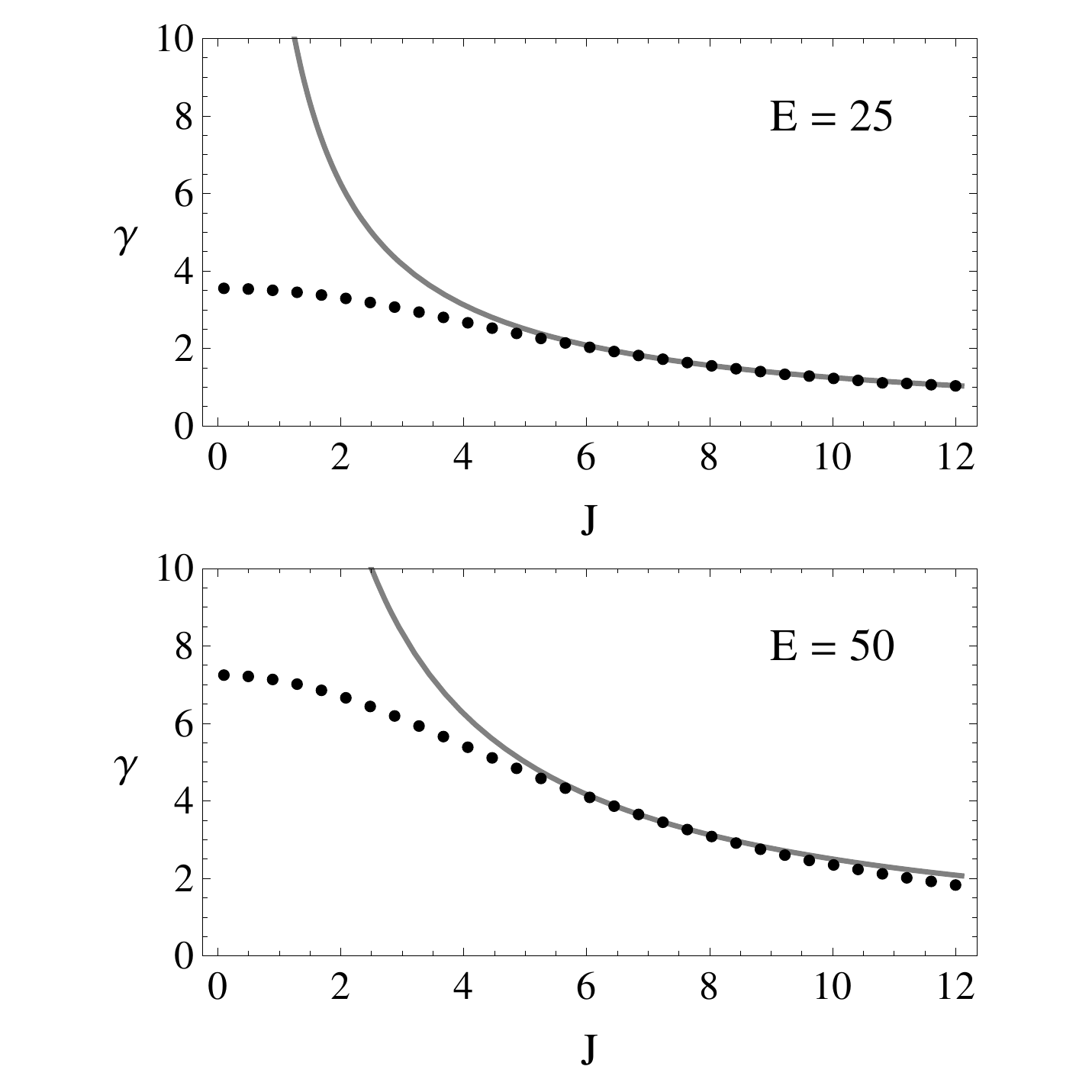}}
\hspace{-1.3cm}
\subfigure[\,Schw \, $\bb = 0 $]{\label{stringFIG_10b}\includegraphics[width=0.4\hsize]{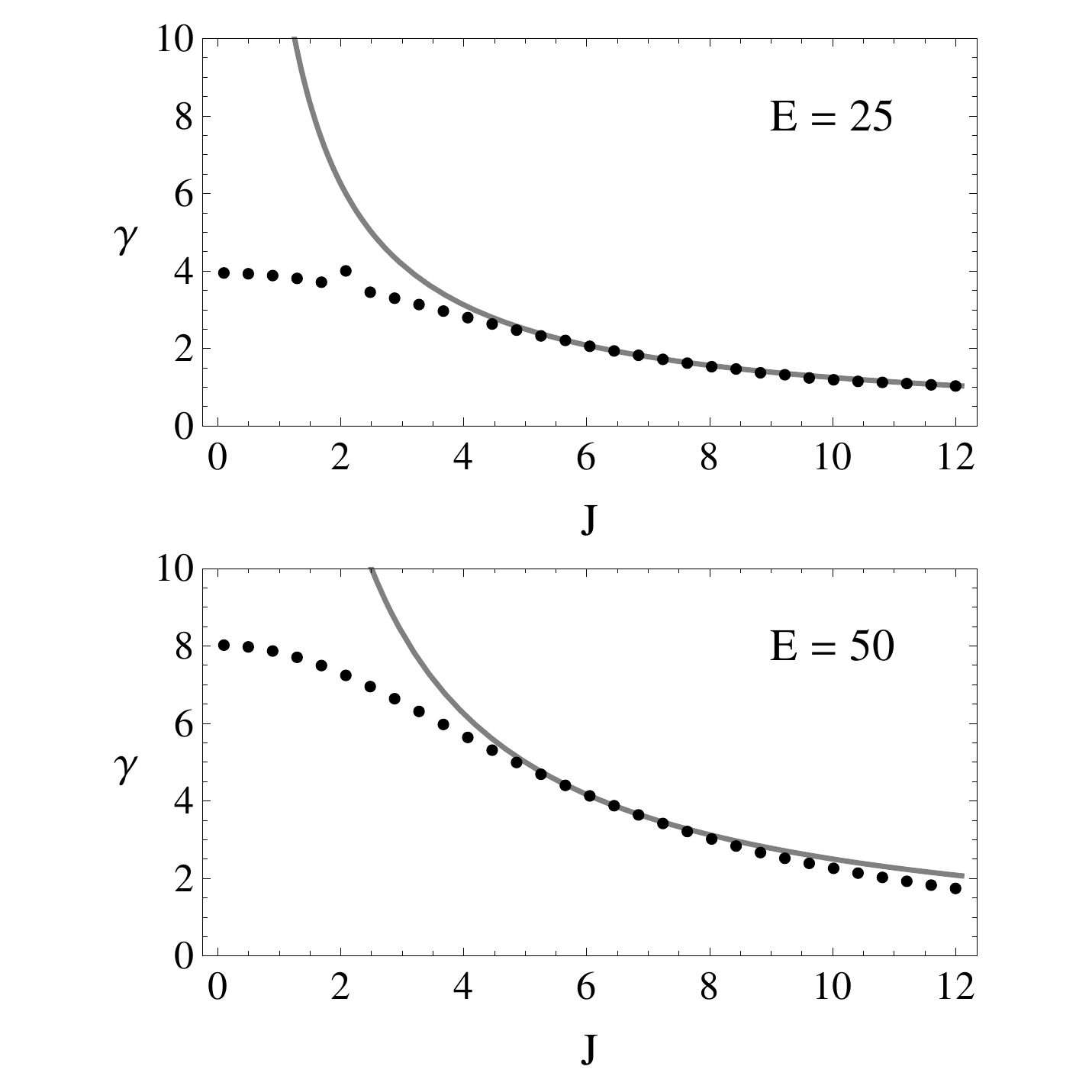}}
\hspace{-1.3cm}
\subfigure[\,RN NS \, $\bb = 1.14$]{\label{stringFIG_10c}\includegraphics[width=0.4\hsize]{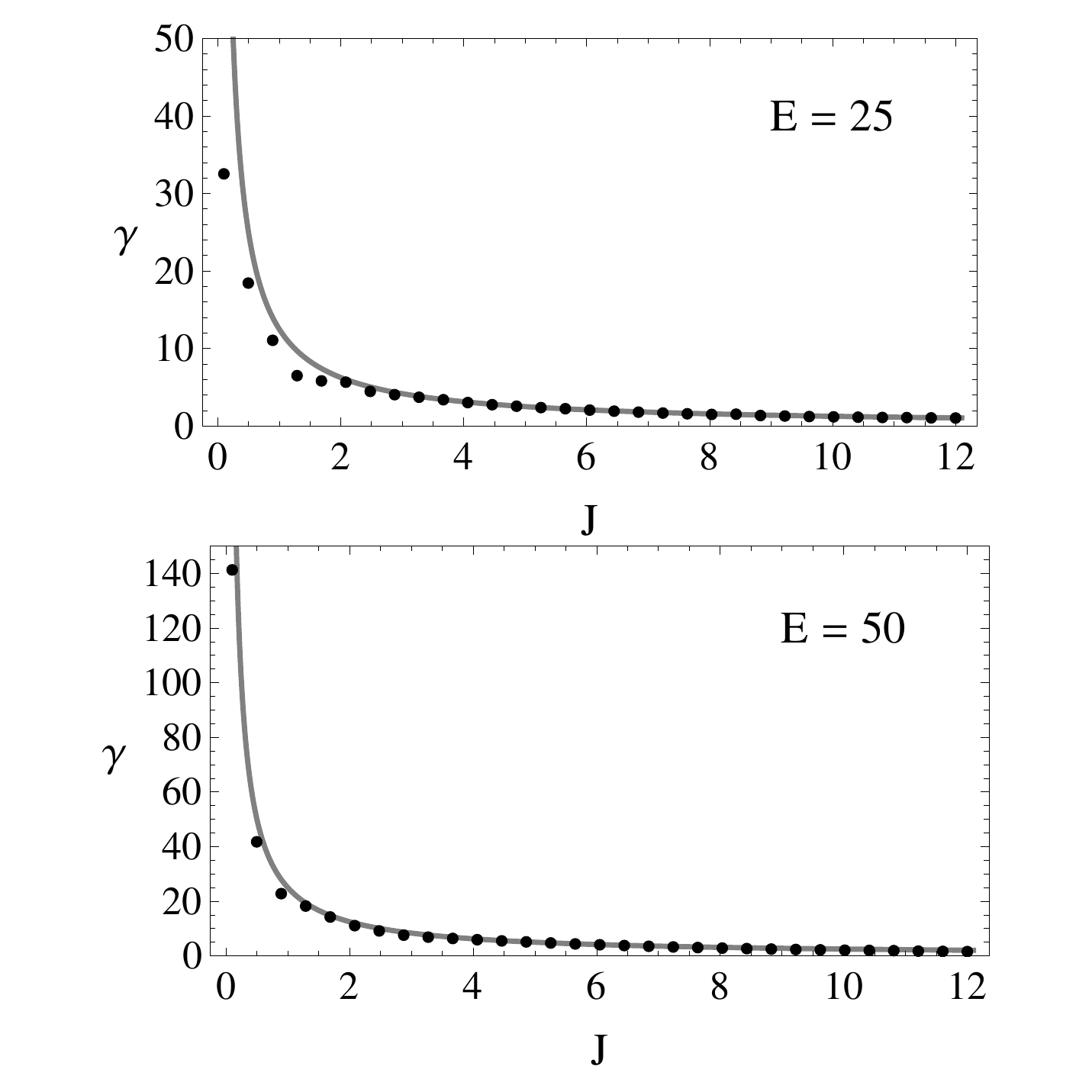}}
\caption{\label{stringFIG_10}
Maximal and extremal (topical) Lorentz gamma factor $\gamma$ in dependence on current $J$. Solid curve shows predicted maximal acceleration $\gamma_{\rm max}$ for given $J$, eq (\ref{gmmax}). The black points are numerically calculated $\gamma_{\rm top}$ from the sample under consideration. We use two energies, $E=25$ and $E=50$. We see clear difference between behavior corresponding to the black hole and naked singularity spacetimes.
} 
\end{figure}

We can see that with the tidal charge of the black holes increasing, efficiency of the transmutation effect increases in both the topical accesible Lorentz factor $\gamma_{\rm top}$ and the relative ratio of strongly accelerated string loops $R_{\gamma( \rm max)/3}$, going up to $\gamma_{\rm top} \sim 4.6$ and $R_{\gamma(\rm max)/3} \sim 0.12$ for near extreme black holes. This behavior can be explained by the fact that as the tidal charge increases, the black hole horizon shrinks and the regions exhibiting increasingly strong gravity are exposed and available for the transmutation process.

For naked singularity spacetimes, the efficiency is higher than in the case of black holes. There is a strong jump up in both $\gamma_{\rm top}$ and $R_{\gamma(\rm max)/3}$ for near extreme naked singularities; a sudden steep decrease of both these quantities occurs at $\bb \sim 2$. Then the topical Lorentz factor slightly increases with tidal charge increasing  at values $\bb > 2$, while the relative ratio of strongly accelerated strings $R_{\gamma(\rm max)/3}$ remains nearly constant with the tidal charge increasing at $\bb>2$, after steep increase at $\bb \sim 2$ to the values relevant at 
$\bb<2$ - see Fig. \ref{stringFIG_9}. 

It is crucial to demonstrate the efficiency of the transmutation effect of the braneworld RN spacetimes by comparing the extremal Lorentz factor $\gamma_{\rm top}$ generated by numerical experiments and reflecting the chaotic character of the string loop motion with the theoretical maximally available Lorentz factor $\gamma_{\rm max}$. We compare the dependences 
$\gamma_{\rm top}(\JJ;\EE,\bb)$ and $\gamma_{\rm max}(\JJ;\EE,\bb)$ for properly selected values of the string loop energy $\EE$ and tidal charge $\bb$. The numerical results are illustrated in Fig. \ref{stringFIG_10}). In the black hole spacetimes, the topical Lorentz factor $\gamma_{\rm top}(\JJ)$ follows the theoretical limit given by $\gamma_{\rm max}(\JJ)$ in a limited range of the angular momentum parameter (down to $\JJ \sim 5$) and for lower values of $\JJ$, the topical factor $\gamma_{\rm top}(\JJ)$ is subtantially lower than the theoretical limit $\gamma_{\rm max}(\JJ)$ and remains finite in the limit of $\JJ \lightarrow 0$. There is $\gamma_{\rm top}(\JJ \lightarrow 0) \sim 4$ for energy parameter $\EE=25$ and $\gamma_{\rm top}(\JJ \lightarrow 0) \sim 8$ for $\EE=50$. We have found only weakly ultrarelativistic string loops with the topical $\gamma$ factor slowly increasing with tidal charge increasing; this slow increase - see Fig. \ref{stringFIG_10a} and Fig. \ref{stringFIG_10b} - is caused by the fact that increasing tidal charge exposes deeper strong gravity regions in vicinity of the event horizon. 

On the other hand, in the naked singularity RN spacetimes of the type NS1 and NS2, the behavior of the topical Lorentz factor $\gamma_{\rm top}(\JJ)$ almost exactly mimics the maximal Lorentz factor $\gamma_{\rm max}(\JJ)$ and can reach strongly ultrarelativistic values; we can easily obtain $\gamma \sim 100$ as demonstrated in Fig. \ref{stringFIG_10c}. In the naked singularity spacetimes with tidal charge close to the extreme value of $\bb=1$, the strong gravity region is rather complex and extremely deep, causing the unexpected extremely strong efficiency of the transmutation phenomena. Our numerical results demonstrate that both high Lorentz factor $\gamma_{\rm top}$ and large relative ratio factor of efficient acceleration $R_{\gamma(\rm max)/3}$ can be obtained in the naked singularity spacetimes with tidal charge in the range  $1<\bb<2$.

The physical reason for such an enormously strong distincion of the trasmutations process efficiency in the black hole and naked singularity spacetimes is related to the non-existence of an event horizon in the naked sigularity spacetimes. The transmutation efficiency is highest at the deepest parts of the gravitational field of black holes or naked singularities. After bouncing at the deepest part of the potential well of a black hole, the string loop crosses the event horizon and is captured by the black hole. On the other hand, there is no such barierre for escaping of the string loops in the field of naked singularities.

The acceleration of the string loops works equally for both cosmic strings that could be remnants of some phase transitions in the early universe \cite{Wit:1985:NuclPhysB:,Vil-She:1994:CSTD:}, and for plasma demonstrating a string-like behavior either due to magnetic field line tubes captured in the plasma\cite{Sem-Dya-Pun:2004:Sci:} or due to thin flux tubes of plasma that can be effectively described as a one-dimensional string \cite{Spr:1981:AA:}. We can expect that in such situations the relevant physics can be described by the string dynamics instead of much more complex magnetohydrodynamics governing plasma in general situations.

In the case of the cosmic strings it could be interesting to speculate about acceleration of cosmic string loops by primordial braneworld naked singularities that could create some traces observable  in the Cosmic Microwave Background - for the role of cosmic strings in the era of recombination see \cite{Pog-Vach:1999:PHYSR4:}. 

The magnetized plasma string-like structures could be relevant in acceleration of collimated jets observed in accreting astrophysical systems ranging from young stars, stellar mass black holes or neutron stars, to supermassive black holes (or, alternatively, braneworld naked singularities) in quasars and active galactic nuclei. As discussed in the pioneering work of Jacobson and Sotiriou \cite{Jac-Sot:2009:PHYSR4:}, processes with "magnetic" strings could arise near equatorial plane of  accreting systems and due to the transmutation process of converting the internal string energy to the kinetic energy of their translational motion a stream of string loops moving along the axis could appear representing thus a well collimated jet. The "magnetic" string loops could be created by magnetized plasma leaving the accretion discs at their inner edge as plasma flux tubes posses both angular momentum  and tension. Such a model needs a detailed study, at least because of the fact that the magnetic tension is not a constant, but depends on the string length; however, the main characteristics of our study have to be relevant \cite{Jac-Sot:2009:PHYSR4:}. 

In their study, Jacobson and Sotiriou \cite{Jac-Sot:2009:PHYSR4:} concluded that only mildly relativistic acceleration of string loops is possible, they reported maximal Lorentz factor $\gamma \sim 1.1$ ($v/c \sim 0.4$) for energy parameter $E=37$, when jets observed in binary systems containing a neutron star \cite{Liv:1999:PhysRep:} can be explained. They claimed applicability of the transmutation process in the Newtonian limit, with non-relativistic velocities, when the process can be relevant for accreting young stars. 

We are able to report larger final velocity of accelerated jets in the black hole spacetimes for the same energy parameter, with Lorentz factor $\gamma \sim 6$ ($v/c \sim 0.93$), which is still mildly relativistic, but applicable also in microquasars, binary systems containing a stellar mass black hole. \footnote{The difference in comparison to the result reported in \cite{Jac-Sot:2009:PHYSR4:} is caused by the fact that we have considered all possible values of the angular momentum parameter $J$.} Further, we have found quite new and interesting possibility to obtain string loops accelerated to ultra-relativistic final velocities, with Lorentz factor $\gamma \sim 100$ for the same energy parameter, in the field of braneworld naked singularities. Such enormous acceleration could be relevant in the most exotic situations related to jets observed in quasars and active galactic nuclei \cite{Liv:1999:PhysRep:}. The ultra-relativistic acceleration is related to the basical properties of naked singularities (deep gravitational potential and non-existence of an event horizon) and has nothing to do with rotation of the objects. We thus can conclude that the string loop transmutation effect can explain the jet production in the whole range of observed phenomena and the rotational effects are not relevant part of these effects, contrary to the Blanford-Znajek effect assumed in the standard electromagnetic model of acceleration of ultra-relativistic jets occuring in quasars and active galactic nuclei. 

For energy and angular momentum parameters of a string loop fixed, a negative (positive) braneworld tidal charge slightly decreases (increases) the energy efficiency of the transmutation process (maximal Lorentz factor) in the braneworld black hole spacetimes in comparison to the Schwarzschild case, but we observe a substantial jump of the energy efficiency after crossing the region of braneworld naked singularities. 
 
%
\section{Regular and chaotic motion of string loops}
%

Motion of strings is generally a chaotic motion. This statement holds even in the case of the motion of axisymmetric string loops in symmetric spacetimes that can be described by a Hamiltonian corresponding, due to the assumed symmetries of the string and the spacetime, to the motion of a particle \cite{Lar:1994:CLAQG:}. Nevertheless, in the motion of the string loops can occur, in analogy with the motion of charged test particles, "islands of regularity". We shall discuss the connection of the regular and chaotic motion of string loops in the RN spacetimes. The relativistic chaos of the charged test particle motion and the transition from regular to chaotic regime of the motion has been studied in \citep{Kop-Kar-Kov-Stu:2010:ASTRJ:,Sem-Suk:2010:MNRAS:}. In both cases of string loops and charged test particles, the regular motion is related to the stable stationary points of the motion.

The stable stationary points, i.e., stable equilibria at $\x^\alpha_0=(\rr_0,\tt_0)$, correspond to the local minima of the energy boundary function 
$\EE_{\rm b}(\rr,\tt,\JJ,\bb)$ that coincide with the critical points of the Hamiltonian introduced by (\ref{HamHam}) and reflecting the string motion in a spherically symmetric spacetime. The Hamiltonian can be rewritten in the form
\beq
 H = U(\rr) {\p}_r^2 + V(\rr) \p_\theta^2 + W(\rr,\tt).
\eeq 
Introducing a small parameter $\epsilon << 1$, we can rescale coordinates and momenta by the relations 
\beq
 \x^\alpha = \x^\alpha_0 + \epsilon \hx^\alpha, \quad \p_\alpha = \epsilon \hp_\alpha,
\eeq
applied for the coordinates $\alpha \in {\rr,\tt}$. We can expand the functions $U(\rr), V(\rr)$ and $W(\rr,\tt)$ into Taylor series and express the Hamiltonian in separated parts according to the power of $\epsilon$
\beq
 H (\hp_\alpha,\hx^\alpha) = H_0 + \epsilon H_1(\hx^\alpha) + \epsilon^2 H_2(\hp_\alpha,\hx^\alpha)
  + \epsilon^3 H_3(\hp_\alpha,\hx^\alpha) + \ldots, \label{expand1} 
\eeq
where $H_k$ is homogeneous part of the Hamiltonian of degree $k$ considered for the momenta $\hp_\alpha$ and coordinates $\hx^\alpha$. Recall that $\p_\alpha$ is quadratic in (\ref{HamHam}) and apears in $H_k$ only for $k \geq 2$.
The first term $H_0$ (just a number) corresponds to the energy at a local minimum $\EE_0$ and can be canceled by rescaling the energy level 
\beq
 \EE \rightarrow \EE - \EE_0. 
\eeq  
For the second term $H_1(\hx^\alpha)$, depending only on the coordinates $\hx^\alpha$, we can put $H_1(\hx^\alpha) =0 $, because we assume the point corresonding to a local minimum.

We can divide (\ref{expand1}) by the factor $\epsilon^2$ (remember $H=0$) expressing the Hamiltonian in the vicinity of the local minimum in the form
\beq
 H =  H_2(\hp_\alpha,\hx^\alpha) + \epsilon H_3(\hp_\alpha,\hx^\alpha) + \ldots
\eeq
If the factor $\epsilon = 0$, we arrive to an integrable Hamiltonian 
\beq
 H =  H_2(\hp_\alpha,\hx^\alpha) = \frac{1}{2} \sum_\alpha \left((\hp_\alpha)^2 + \omega^2_\alpha (\hx^\alpha)^2 \right)
\eeq
representing a harmonic oscillatory motion. 
The integrable Hamiltonian implies the string motion to be regular (a periodic motion). 
For the string loop motion represented by a point with coordinates $\rr = \rr_0 + \dr,\tt = \tt_0 + \dt$ we obtain the  periodic oscillations determined by the equations 
\beq
 \ddot{\dr} + \omega^2_{\rr} \, \dr = 0, \quad \ddot{\dt} + \omega^2_{\tt} \, \dt = 0,
\eeq
where the locally measured frequencies of the oscillatory motion are given by 
\beq
 \omega^2_\rr  = \frac{\partial^2 \Veff}{\partial \rr^2},\quad \omega^2_\tt  = \frac{1}{r^2 A} \frac{\partial^2 \Veff}{\partial \tt^2}.
\eeq
For local observers in the braneworld \RN{} spacetimes the frequencies of the oscillatory motion around a given stationary point at radius $\rr$ read
\bea
 \omega^2_\rr &=& \frac{4 (b+r^2-2r) \left(4 b r-3 b+r^3-5 r^2+3 r\right)}{r (2 b+(r-3) r)^2},\\
 \omega^2_\tt &=& \frac{4 (r-b) (b+(r-2) r)}{(2 b+(r-3) r)^2}.
\eea

\begin{figure}[t]
\subfigure[$\,\, \EE = 19.8 $]{\includegraphics[width=0.23\hsize]{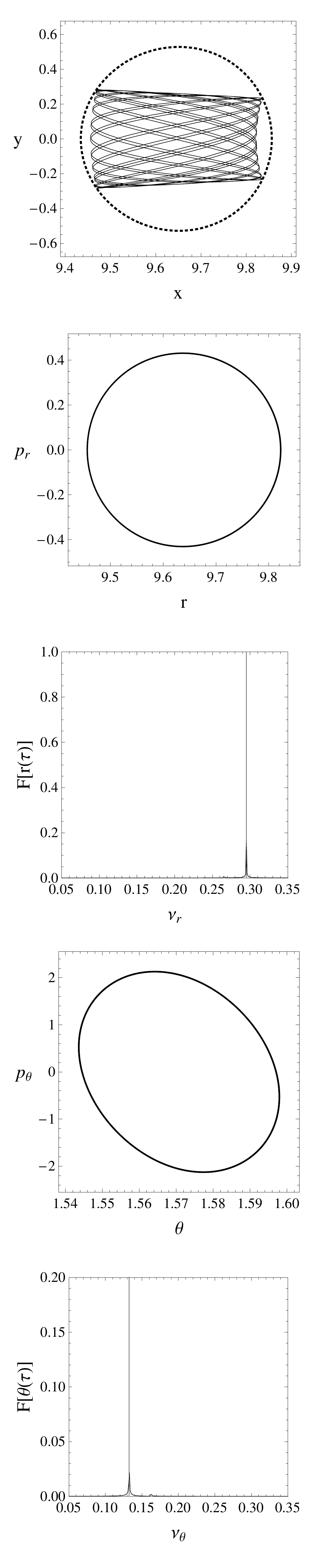}}
\subfigure[$\,\, \EE = 20 $]{\includegraphics[width=0.23\hsize]{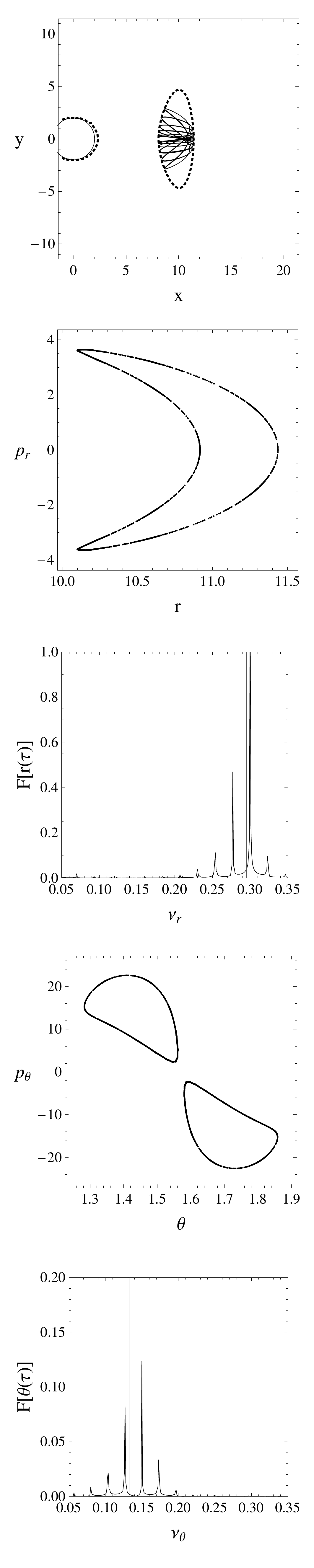}}
\subfigure[$\,\, \EE = 20.15 $]{\includegraphics[width=0.23\hsize]{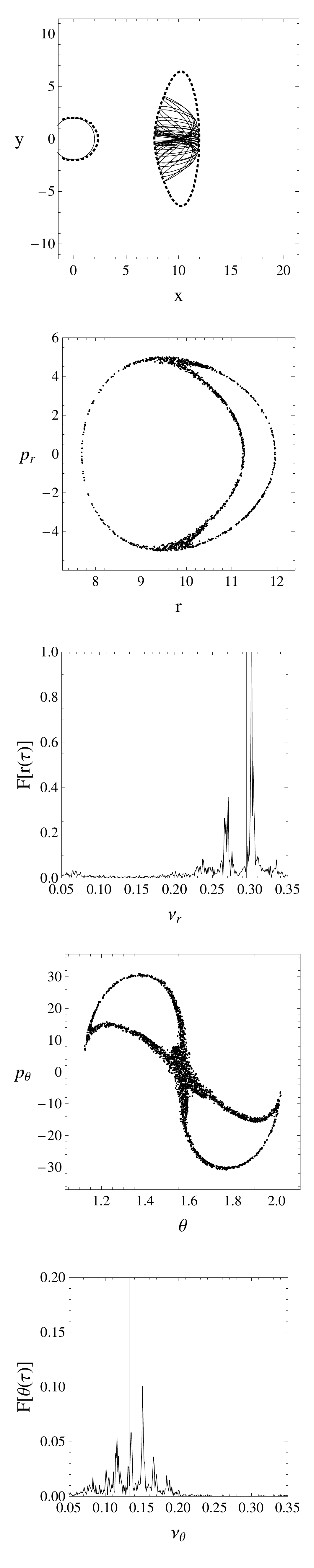}}
\subfigure[$\,\, \EE = 20.5   $]{\includegraphics[width=0.23\hsize]{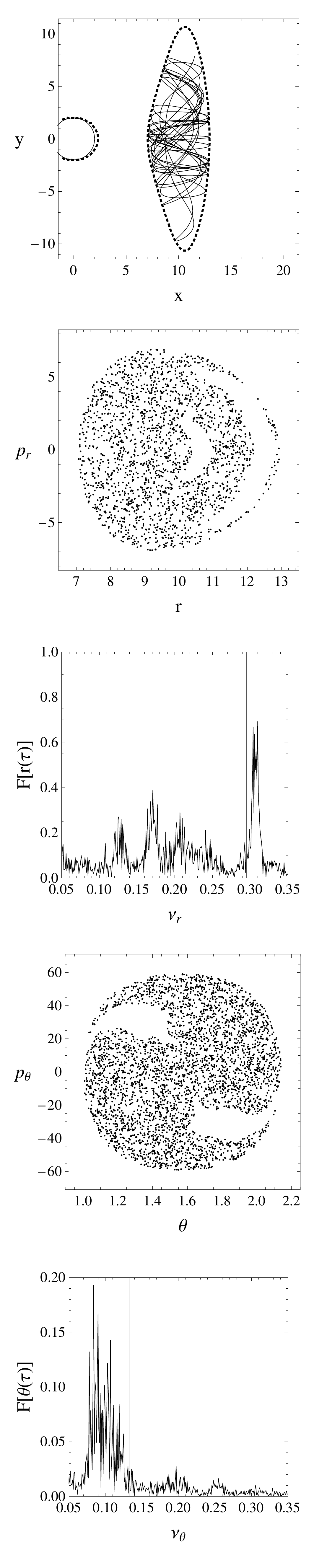}}
\vspace{-0.2cm}
\caption{ \label{stringFIG_11}
Transition from the regular to the chaotic regime of the string loop motion in the Schwarzschild black hole spacetime with $\bb=0$. The string loop is starting from the rest near the local minimum located (for the string angular momentum $\JJ=11$) at $r_0\doteq 9.64,\theta_0 = \pi/2$, with successively increasing energy $\EE$. For every energy level we plotted the string loop trajectory, the Poincare surface sections ($\rr,\p_r$),($\tt,\p_\tt$) and the Fourier spectrum for both coordinates $\rr$ and $\theta$ \citep{Ott:1993:book:,Cont:2002:book:,Reg:2006:book:}. The vertical lines in the Fourier spectra are the frequencies $ \nu_\rr = \omega_r / (2 \pi), \nu_\tt = \omega_\tt / (2 \pi)$.
For small energies the motion is regular, above the energy $\EE \sim 20.15$, the chaotic regime begins.
} 
\end{figure}
\begin{figure}[t]
\subfigure[$\,\, \EE = 15.299 $]{\includegraphics[width=0.23\hsize]{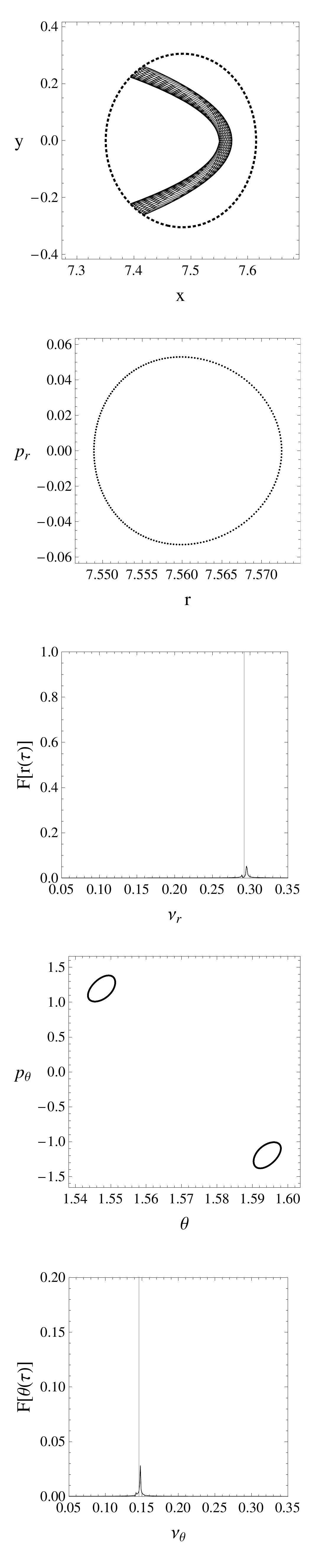}}
\subfigure[$\,\, \EE = 15.45 $]{\includegraphics[width=0.23\hsize]{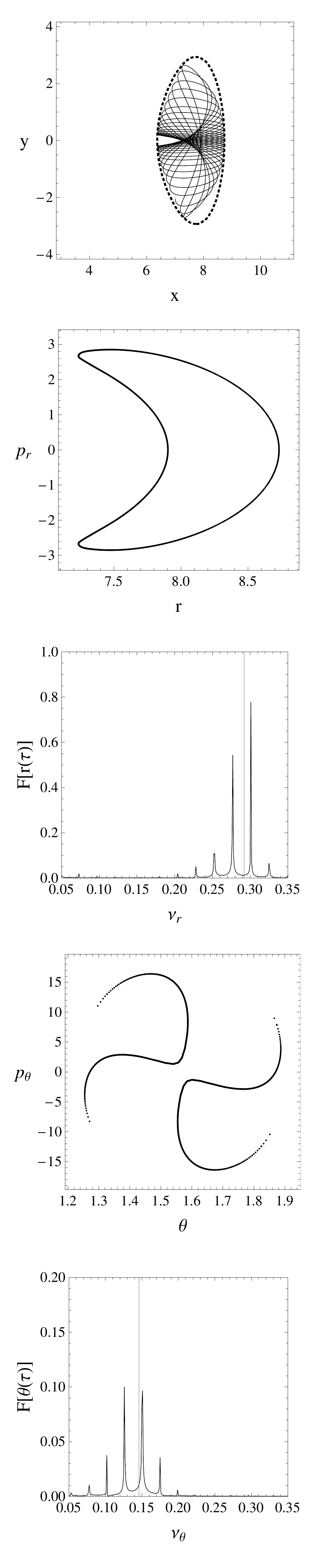}}
\subfigure[$\,\, \EE = 15.5 $]{\includegraphics[width=0.23\hsize]{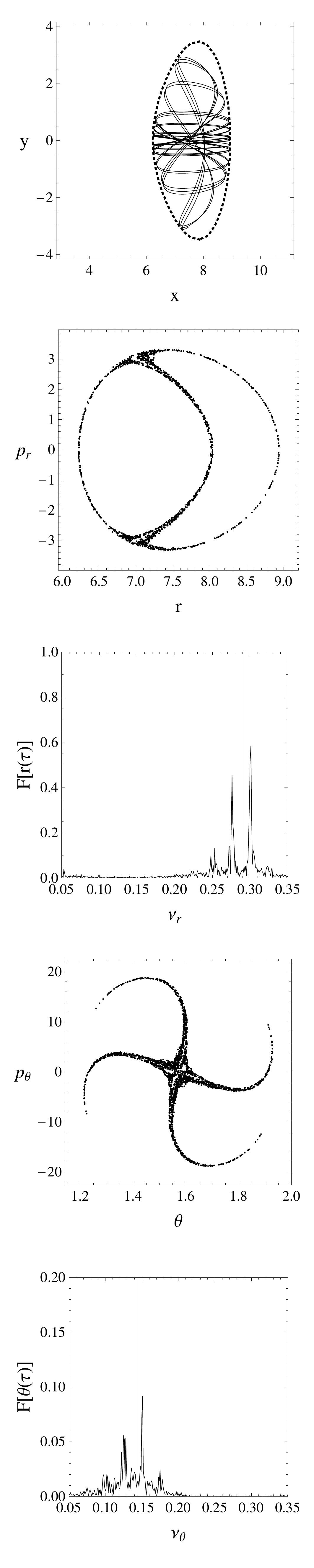}}
\subfigure[$\,\, \EE = 15.9   $]{\includegraphics[width=0.23\hsize]{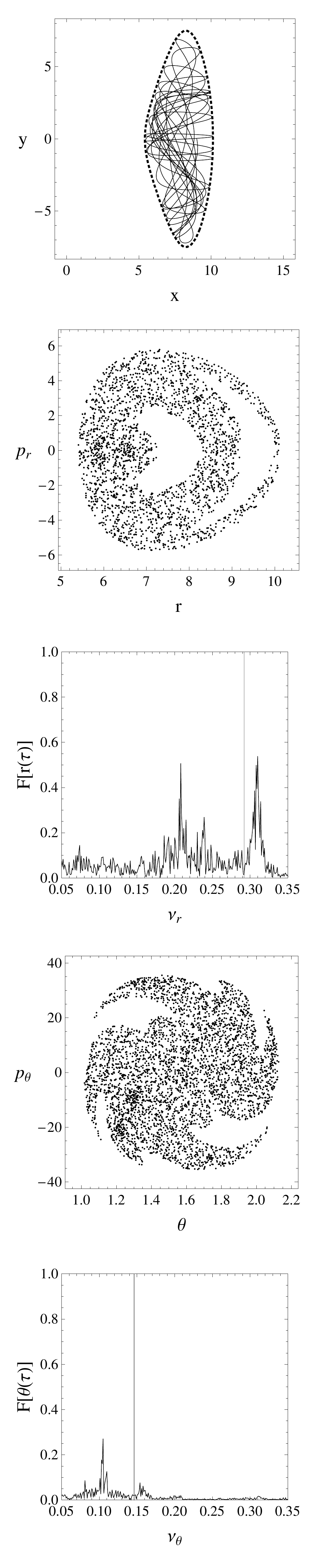}}
\vspace{-0.2cm}
\caption{ \label{stringFIG_12}
Transition from the regular to the chaotic regime of the string loop motion in the RN naked singularity spacetime with $\bb=1.1$. The string loop is starting form the rest near the local minimum located (for the string angular momentum $\JJ \doteq 8.72$) at $r_0\doteq 7.48,\theta_0 = \pi/2$ with succesively increasing energy $\EE$. For every energy level we plotted the string loop trajectory, the Poincare surface sections ($\rr,\p_r$),($\tt,\p_\tt$) and the Fourier spectrum for both coordinates $\rr$ and $\theta$ \citep{Ott:1993:book:,Cont:2002:book:,Reg:2006:book:}. The vertical lines in the Fourier spectra has ratio $ \nu_{\rr}:\nu_\tt = 2:1$.
For small energies the motion is regular, above the energy $\EE \sim 15.5$, the chaotic regime begins.
} 
\end{figure}

\begin{figure*}
\includegraphics[width=\hsize]{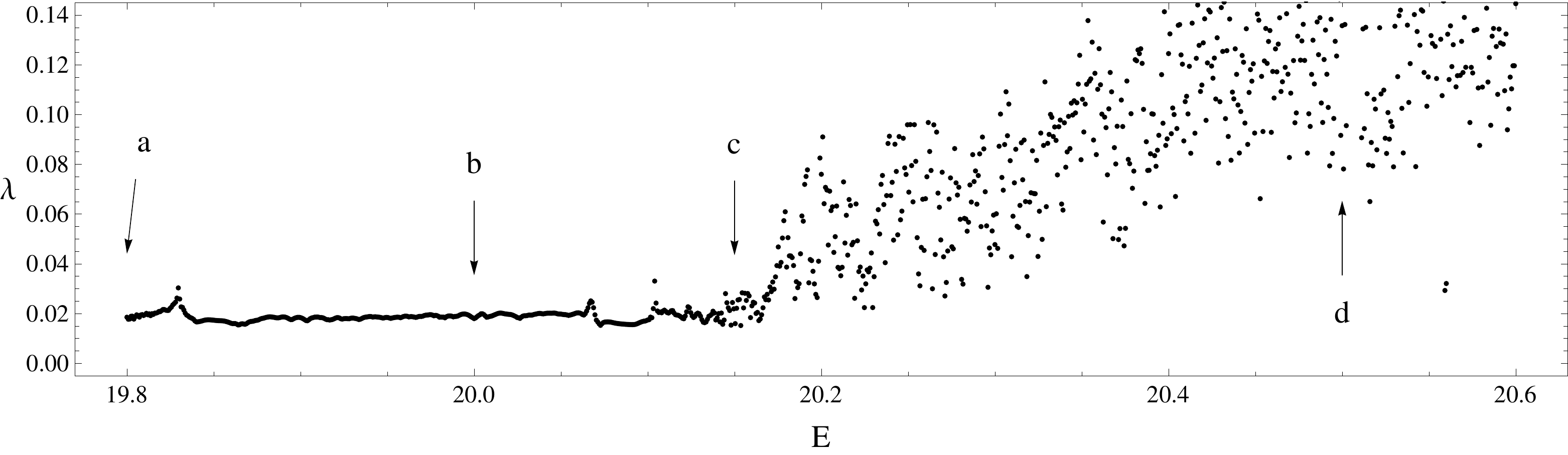}
\caption{ \label{stringFIG_13}
Evolution of the maximal Lyapunov exponent in dependence of on the string loop energy, related to Fig. \ref{stringFIG_11}. For small energies the motion is regular, for bigger energies the motion is chaotic - this is manifestation of the KAM theorem. The transition between the regular/chaotic regimes occurs approximately at $\EE \sim 20.15$. Letters denote the individual cases in Fig. \ref{stringFIG_11}.
}
\end{figure*}

According to the Kolmogorov-Arnold-Moser (KAM) theory \citep{Kol:1954:DAN:,Arn:1963:RMS:,Mos:1962:NAWG:}, a string loop  will oscillate in a quasi-periodic motion, if the parameter $\epsilon$ remains small. As the parameter $\epsilon$ grows, the condition $\epsilon << 1$ becomes violated, the nonlinear parts in the Hamiltonian become stronger, and the string loop enters the nonlinear, chaotic regime of its motion. Increase of non-linearity of a system moving in vicinity of its local stable equlibrium point is caused by increase of its energy. The situation demonstrating succesive transfer from purely regular, periodic motion through quasi-periodic motion to purely chaotic motion of a string loop is illustrated in Fig. \ref{stringFIG_11} for a black hole spacetime, and in Fig. \ref{stringFIG_12} for a naked singularity spacetime. In the second case, we have chosen a special ratio of the oscillatory frequencies with $\omega_\rr : \omega_\tt = 2 : 1$, in order to demonstrate a very simple regular motion. The Poincare surface sections in the phase space and the Fourier transforms of the oscillatory motion in the radial and latitudinal direction clearly represent the transfer to the chaotic motion. Of course, in the entering phase of the motion with lowest energy, the string loop motion is fully regular and periodic and is represented by appropriate Lissajouse figures. We see that the transition from the regular to chaotic motion is of the same character for both black holes and naked singularities. 

It is convenient to represent the transfer to the chaotic system by an appropriate Lyapunov coefficient. The chaotic systems are sensitive to initial conditions and we can follow two string loop trajectories separated at the initial time $t_0$ by a small phase-space distance $d_0$. As the system evolves due to the increasing energy parameter, the two orbits will be separated at an exponential rate if the motion of the string loops enters the chaotic regime. The Lyapunov exponent \citep{Ott:1993:book:}
\beq
\lambda_{\rm L} = \lim_{d_0 \rightarrow 0 \atop t  \rightarrow \infty } \left( \frac{1}{t} \ln \left( \frac{d(t)}{d_0} \right) \right)
\eeq
is describing the two orbits separation and hence the measure of chaos. The transition from the regular to the chaotic regime of the string loop motion is clearly visible due to the evolution of the maximal Lyapunov exponent \cite{Ben-etal:1980:LCE1:} demonstrated in Fig. \ref{stringFIG_13}. We clearly see increasing measure of chaos with increasing energy of the moving string loop when the energy limit of the regular or quasi-periodic motion is crossed. Notice that the transition from the regular to the chaotic regime of the motion can be the solution to the "focusing" problem of the string trajectories discussed in \cite{Jac-Sot:2009:PHYSR4:}.

The quasi-periodic character of the motion of string loops trapped in a toroidal space around the equatorial plane of a black hole (naked singularity) suggests another interesting astrophysical application, related to the high-frequency quasi-periodic oscillations (HF QPOs) of X-ray brightness that had been observed in many Galactic Low Mass X-Ray Binaries containing neutron~stars \citep[see, e.g.,][]{Bar-Oli-Mil:2005:MONNR:,Bel-Men-Hom:2007:MONNR:BriNSQPOCor} or black holes \citep[see, e.g.,][]{Rem:2005:ASTRN:,Rem-McCli:2006:ARASTRA:}. Some of the~HF~QPOs come in pairs of the~upper and lower frequencies ($\nu_{\mathrm{U}}$, $\nu_{\mathrm{L}}$) of {\it twin peaks} in the~Fourier power spectra. Since the~peaks of high frequencies are close to the~orbital frequency of the~marginally stable circular orbit representing the~inner edge of Keplerian discs orbiting black holes (or neutron~stars), the~strong gravity effects must be relevant in explaining HF~QPOs \citep{Tor-etal:2005:ASTRA:}. Usually, the Keplerian orbital and epicyclic (radial and latitudinal) frequencies of geodetical circular motion are assumed in models explaining the HF QPOs in both black hole and neutron star systems. However, neither of the models is able to explain the HF QPOs in all the microquasars \cite{Tor-etal:2011:ASTRA:}. Therefore, it is of some relevance to let the string loop oscillations, characterized by their radial and vertical (latitudinal) frequencies, to enter the game, as these frequencies are comparable to the epicyclic geodetical frequencies, but slightly different, enabling thus some corrections to the predictions of the models based on the geodetical epicyclic frequencies. Of course the frequencies of the string loop oscillations in physical units have to be related to distant observers when they take the form 
\bea
 \omega^2_{\rm r} &=& \left( \frac{c^3}{GM} \right)^2 \, \frac{r^3 -5r^2 + 3r -3b + 4b r}{r^5}\\
 \omega^2_{\rm \theta} &=& \left( \frac{c^3}{GM} \right)^2 \, \frac{r-b}{r^4}.
\eea
It is quite interesting that the latitudinal oscillatory frequency of the string loop equals to the latitudinal frequency of the epicyclic geodetical motion as observed by distant observers - for details see \cite{Stu-Kot:2009:GRG:}. The radial frequencies of the string loop oscillations and the geodetical epicyclic motion are, however, different, enabling thus some substantial changes of the relation of the frequency ratio for HF QPOs modelled by the string loop motion and the geodetical epicyclic oscillations. In the black hole spacetimes that can be used to describe the external field of neutron stars, the most crucial difference is given by the equality of frequencies of the radial and latitudinal oscillatory motion of string loops in vicinity of the horizon, but above the radius of the photon circular orbit. Such an effect enables application of the string loop oscillatory motion to explanation of the HF QPOs in neutron star systems, where the model with both epicyclic geodetical oscillations is excluded since their frequencies cannot be equal in the black hole spacetimes - for details see \cite{Urb-etal:2010:ASTRA:}.  We plan to study the HF QPO phenomena in a future paper, related also to investigations of the properties of the transfer between the quasi-periodic and chaotic regime of the motion. 

%
\section{Conclusions}
%

Motion of current (angular momentum) carrying axisymmetric string loops in spherically symmetric or axisymmetric spacetimes can be described by a point particle Hamiltonian containing a special term reflecting the interplay of the string tension and angular momentum that is the cause of the chaotic character of the motion. The Hamilton equations then imply a simple effective potential determining the string loop motion. We have studied character of the effective potential of the axisymmetric string loop motion, transition from the regular to chaotic motion in vicinity of stable equlibrium points of the motion, and possible acceleration of string loops due to the transmutation effect occuring in the strong gravitational field of braneworld \RN{} black holes and naked singularities. Our results are relevant also for the motion of electrically uncharged string loops in the field of \RN{} black holes and naked singularities carrying an electric charge $Q$ when we can use the transposition $\bb \lightarrow Q^2$. For electrically charged string loops the situation is more complex, but still can be treated by using an appropriately chosen Hamiltonian \cite{Lar:1993:CLAQG:} - we plan to realize such a study in a future work. In the MHD string approximation \citep{Sem-Ber:1990:ASS:,Spr:1981:AA:} the string transmutation effect in vicinity of a black hole horizon or naked singularity could serve as a toy model of acceleration of jets in quasars (AGN) or microquasars, and is clearly of potentially high astrophysical relevance \cite{Jac-Sot:2009:PHYSR4:,Stu-Kol:2012:PHYSR4:}. 

Our results can be summarized as follows

\begin{itemize}

\item
In the braneworld \RN{} black hole spacetimes, the motion of string loops is of the same character as in the Schwarzschild spacetime, discussed in \cite{Kol-Stu:2010:PHYSR4:}. The tidal charge causes quantitative differences only. The region accesible for the string motion is shifted to larger distances from the black hole horizon as the tidal charge decreases. Therefore, the string moves in relatively shallow gravitational potential of the black hole with negative tidal charges, while the potential becomes much deeper around the horizon of black holes with positive tidal charge. The stable equilibrium points of the string loops corresponding to local minima of the effective potential occur only in the equatorial plane of the system of the black hole and axisymmetric string loop and the string loop motion can cross the equatorial plane. 

\item
In the field of braneworld \RN{} naked singularities the situation is qualitatively different. Contrary to the motion in the black hole spacetimes, capturing of the string loop by the physical singularity is excluded in the naked singularity spacetimes - only trapped states of string loops and string loops escaping to infinity are allowed. Moreover, a new, off-equatorial minima of the effective potential emerge, besides the standard equatorial local minima, in the naked singularity spacetimes. For some special values of the angular momentum and energy parameters of the string loops, their motion across the equatorial plane is forbidded -- see Fig. \ref{stringFIG_7}.

Notice that for the charged particle motion in the \RN{} spacetimes the off-equatorial minima of the effective potential do not exist due to the spherical symmetry, but in the case of the motion of charged particle in the field of \KN{} naked singularities such off-equatorial minima can exist \citep{Cla-Fel:1982:NUCIBS:,Stu-Hle:1998:APS:,Kov-Stu-Kar:2008:CLAQG:}.
In the case of the motion of the string loops, again, naked singularity is necessary in order to obtain the off-equatorial equilibrium points. 

\item
The transition from the regular motion around a stable equlibrium state of a string loop to the chaotic motion has the same character for both the braneworld RN black holes and naked singularities. The character of the regular motion is represented by the Lissajousse figures that become especially simple if the frequency ratio of the oscillations in the radial and latitudinal directions is comensurable with small integers. The transition reflects in an illustrative way the KAM theorem of the general quasiperiodic motion. 

\item
String loop acceleration to ultrarelativistic velocities (string transmutation effect) is stronger and occurs more frequently in the field of braneworld \RN{} naked singularities than black holes - see Figs. \ref{stringFIG_9}, \ref{stringFIG_10}. In the naked singularity spacetimes, highly ultrarelativistic Lorentz factors ($\gamma \sim 100$) are easily attainable and are comparable to the theoretical limit of $\gamma_{\rm max}$ even for very low angular momentum parameter of the string loop, contrary to the black hole case where the extremal Lorentz factor $\gamma_{\rm top}$ is by more than one order smaller for similar small angular momentum parameters of the string loops undergoing the transmutation process, being strongly supressed in comparison with theoretical maximum $\gamma_{\rm max}$.

\end{itemize}

We can conclude that there is a variety of astrophysically relevant phenomena that could give a strong signature of the presence of the braneworld naked singularity spherically symmetric spacetimes in the properties of the string loop motion. For braneworld black holes we can obtain only quantitative signatures of the presence of the tidal charge parameter. In all the braneworld black hole or naked singularity spherically symmetric spacetimes we can expect observationally relevant signatures of the presence of the hidden dimension reflected by the tidal charge.

\section*{Acknowledgements}
The authors would like to express their acknowledgements for the Institutional support of the Faculty of Philosophy and Science of the Silesian University at Opava, the internal student grant of the Silesian University SGS/2/2010 and EU grant Synergy CZ.1.07/2.3.00/20.0071



\end{document}